\documentclass[apj]{emulateapj}

\usepackage{graphicx,ifthen,url,float,lscape,color,ulem}
\usepackage{natbib}

\newcommand {\kms}{km s$^{-1}$}

\def\ltsima{$\; \buildrel < \over \sim \;$}
\def\simlt{\lower.5ex\hbox{\ltsima}}
\def\gtsima{$\; \buildrel > \over \sim \;$}
\def\simgt{\lower.5ex\hbox{\gtsima}}

\def\ts     {\thinspace}
\def\kms    {\ifmmode{{\rm \ts km\ts s}^{-1}}\else{\ts km\ts s$^{-1}$}\fi}
\def\msol   {\ifmmode{{\rm M}_{\odot}}\else{M$_{\odot}$}\fi}
\def\lsol   {\ifmmode{{\rm L}_{\odot}}\else{L$_{\odot}$}\fi}
\def\zsol   {\ifmmode{{\rm Z}_{\odot}}\else{Z$_{\odot}$}\fi}

\def\ci     {\ifmmode{{\rm C}{\rm \small I}}\else{C\ts {\scriptsize I}}\fi}
\def\hi     {\ifmmode{{\rm H}{\rm \small I}}\else{H\ts {\scriptsize I}}\fi}
\def\hh     {\ifmmode{{\rm H}_2}\else{H$_2$}\fi}
\def\cone {\ifmmode{{\rm C}{\rm \small I}(^3\!P_1\!\to^3\!P_0)}
     \else{C\ts {\scriptsize I}{\small$(^3\!P_1\!\to\,^3\!P_0)$}}\fi}
\def\ctwo {\ifmmode{{\rm C}{\rm \small I}(^3\!P_2\!\to\,^3\!P_1)}
     \else{C\ts {\scriptsize I}{\small$(^3\!P_2\!\to\,^3\!P_1)$}}\fi}
\def\cij {\ifmmode{{\rm C}{\rm \small I}\,(^3P_i\to^3P_j)}\else{C\ts {\scriptsize I}\,{\small$(^3P_i\to^3P_j)$}}\fi}
\def\cii    {\ifmmode{{\rm C}{\rm \small II}}\else{C\ts {\scriptsize II}}\fi}
\def\tex {\ifmmode{{T}_{\rm ex}}\else{$T_{\rm ex}$}\fi}
\def\tmb {\ifmmode{{T}_{\rm mb}}\else{$T_{\rm mb}$}\fi}
\def\tkin {\ifmmode{{T}_{\rm kin}}\else{$T_{\rm kin}$}\fi}
\def\microns {\ifmmode{\mu{\rm m}}\else{$\mu$m}\fi}
\def\nhh   {\ifmmode{n({\rm H}_2)}\else{$n$(H$_2$)}\fi}

\shorttitle{Continuum Overdensities around High-z Quasars}
\begin{document}

\title{No Evidence for Millimeter Continuum Source Overdensities \\
    in the Environments of $z\gtrsim 6$ Quasars}
    
\author{Jaclyn B. Champagne\altaffilmark{1}}
\author{Roberto Decarli\altaffilmark{2,3}}
\author{Caitlin M. Casey\altaffilmark{1}}
\author{Bram Venemans\altaffilmark{2}}
\author{Eduardo Ba\~nados\altaffilmark{4, 5}}
\author{Fabian Walter\altaffilmark{2}}
\author{Frank Bertoldi\altaffilmark{6}}
\author{Xiaohui Fan\altaffilmark{7}}
\author{Emanuele Paolo Farina\altaffilmark{8}}
\author{Chiara Mazzucchelli\altaffilmark{2}}
\author{Dominik A. Riechers\altaffilmark{9}}
\author{Michael A. Strauss\altaffilmark{10}}
\author{Ran Wang\altaffilmark{11}}
\author{Yujin Yang\altaffilmark{12}}

\affil{1. University of Texas at Austin, 2515 Speedway Blvd Stop C1400, Austin, TX 78712, USA}
\affil{2. Max Planck Institute for Astronomy, 
    Heidelberg, Germany 69117}
\affil{3. INAF - Osservatorio di Astrofisica e Scienza dello Spazio di Bologna, via Gobetti 93/3, 40129 Bologna, Italy}
\affil{4. Carnegie-Princeton Fellow}
\affil{5. The Observatories of the Carnegie Institution for Science, 813 Santa Barbara St., Pasadena, CA, 91101, USA}
\affil{6. Argelander Institute for Astronomy, University of Bonn, Auf dem H\"ugel 71, 53121 Bonn, Germany}
\affil{7. Steward Observatory, University of Arizona, Tuscon, AZ, 85721, USA}
\affil{8. Department of Physics, University of California, Santa Barbara, CA, 93106, USA}
\affil{9. Cornell University, 220 Space Sciences Builing, Ithaca, NY, 14853, USA}
\affil{10. Department of Astrophysical Sciences, Princeton University, Princeton, NJ 08544 USA}
\affil{11. Kavli Institute of Astronomy and Astrophysics at Peking University, No. 5 Yiheyuan Road, Haidian District, Beijing, 100871, China}
\affil{12. Korea Astronomy and Space Science Institute, 776 Daedeokdae-ro, Yuseong-gu, Daejeon 34055, Republic of Korea}

\begin{abstract}
Bright high-redshift quasars ($z>6$), hosting supermassive black holes ($\rm M_{BH} > 10^8 M_{\odot}$), are expected to reside in massive host galaxies embedded within some of the earliest and most massive galaxy overdensities. We analyze 1.2\,mm ALMA dust continuum maps of 35 bright quasars at $6<z<7$ and search the primary beam for excess dust continuum emission from sources with L$_{\rm IR} \simgt{ 10^{12}}$\,L$_{\odot}$ as evidence for early protoclusters. We compare the detection rates of continuum sources at $\geq5\sigma$ significance in the fields surrounding the quasars (A$_{\rm eff}$ = 4.3\,arcmin$^2$) with millimeter number counts in blank field surveys.  We discover 15 millimeter sources in the fields excluding the quasars themselves, corresponding to an overdensity $\delta_{\rm gal} \equiv (N_{\rm gal} - N_{\rm exp})/N_{\rm exp} = -0.07\pm0.56$, consistent with no detected overdensity of dusty galaxies within 140 physical kpc of the quasars. However, the apparent lack of continuum overdensity does not negate the hypothesis that quasars live in overdense environments, as evidenced by strong [CII] overdensities found on the same scales to similarly-selected quasars. The small field of view of ALMA could miss a true overdensity if it exists on scales larger than 1\,cMpc, if the quasar is not centered in the overdensity, or if quasar feedback plays a role close to the quasar, but it is most likely that the large line of sight volume probed by a continuum survey will wash out a true overdensity signal. We discuss the necessary factors in determining the bias with which dusty star-forming galaxies trace true dark matter overdensities in order to improve upon overdensity searches in dust continuum.

\end{abstract}

\section{Introduction}

The processes by which dusty star-forming galaxies within large scale structure collapse and evolve over time are unconstrained both in theory \citep{Narayanan2015a, Chiang2017a} and observations \citep{Chapman2009a, Casey2016a}. Cosmological simulations in a cold dark matter Universe show that structures form hierarchically, i.e., that the earliest structures should be less massive than the galaxy clusters and superclusters we see today \citep{Springel2005a}. However, cosmic downsizing \citep{Cowie1986a} suggests that the most massive structures in overdense environments assembled their mass earlier than galaxies in the surrounding field, thus making high-redshift observations critical for studying the progenitors of galaxy clusters seen in the present Universe. It remains observationally unclear when in cosmic history the seeds of these clusters took root and what the evolutionary processes were therein. 

The highest-redshift quasars ($z>6$) host supermassive black holes, where $\rm M_{BH} \gtrsim 10^8 M_{\odot}$ \citep[e.g.,][]{De-Rosa2011a, Mazzucchelli2017b}; theoretical models suggest that they reside in massive host galaxies that are located in some of the highest overdensities at these early cosmic epochs \citep{Overzier2009a, Costa2014a, Sijacki2015a} and grow rapidly at these times through super-Eddington accretion \citep{Madau2014a, Pezzulli2017a}. Galaxy clusters seen at present times are dominated by massive, old elliptical galaxies \citep{Lewis2002a, Skibba2009a} which must have assembled much faster than those in the field. If indeed these massive galaxies coalesced more quickly, we would expect to detect their high-redshift progenitors as highly dust-obscured sources in regions of massive overdensities. 

Thousands of quasi-virialized \footnote{The most massive galaxy clusters are closest to virialization, but generally $z=0$ clusters are still not fully virialized beyond the core; see \citet{Xu2000a}.} galaxy clusters, marked by spatially extended X-ray bremsstrahlung and millimeter Sunyaev-Zel'dovich signatures of a hot intracluster medium (ICM), have been confirmed in the $z<1.5$ Universe \citep{Allen2011a,Kravtsov2012a}. But protoclusters that have not yet collapsed do not exhibit strong ICM emission or absorption, and have typically been identified via overdensities of star-forming galaxies. For instance, evidence for massive overdensities and protoclusters at $z>2$ has been found through Lyman-break galaxies \citep[LBGs;][]{Steidel1998a, Steidel2005a, Capak2011a, Riechers2014a}, Lyman-$\alpha$ emitters (LAEs) via narrow-band imaging \citep{Venemans2002a, Venemans2005a, Toshikawa2017a, Badescu2017A}, and dusty star-forming galaxies via cold dust emission \citep{Chapman2009a, Chiang2013a, Casey2015a, Casey2016a, Hung2016a}. %Wide-field narrowband imaging of LAEs has been used to trace a $z=2.3$ protocluster \citep{Badescu2017A}. 
The FORS2 instrument on the Very Large Telescope \citep[VLT;][]{Appenzeller1998a} as well as Suprime-Cam on the Subaru telescope \citep{Miyazaki2002a} have been instrumental in the detection and imaging of LAE overdensities at $z>4$ \citep[e.g.][]{Shimasaku2003a, Venemans2005a, Kuiper2011a}.  

Similar searches have been performed looking for UV-selected galaxies at higher redshifts, specifically around $z\geq5$ quasars, using both narrow-band imaging \citep[e.g.][]{Banados2013a, Goto2017a, Mazzucchelli2017a, Ota2018a} and spectroscopy \citep[e.g.][]{Farina2017a}. Yet many of them have yielded inconclusive results about the true nature of quasar environments: some searches for LAEs as proof of clustering around quasars have revealed a puzzling lack of overdensities. This has opened questions about the field of view and depth necessary to probe real overdensities \citep[see also][]{Stiavelli2005a}, but one should note that narrow band searches can be particularly hampered by low sensitivity as well as the uncertainty in the redshift of the quasar, which could lead to companions falling out of the narrow band entirely; \citet{Farina2017a} finds a LAE at very close proximity to quasar J0305-3150 using the Multi Unit Spectroscopic Explorer (MUSE) on the VLT, which would have been missed by narrow band searches due to sensitivity constraints. 

The advantages of Ly$\alpha$ in particular are that it is selected in a narrow redshift range in narrow-band imaging, and it can presently be detected out to $z\sim7$, so it can be used as a reliable redshift measure in the earliest massive overdensities \citep[e.g.][]{Djorgovski1985a, Palunas2004a}. At the highest redshifts ($z>7$), single detections of LAEs with large equivalent widths may be indications of ionizing regions around protoclusters \citep{Larson2017a}. However, individual Ly$\alpha$ studies are observationally expensive, as optical observations of LAEs in dense environments becomes challenging due to increasing absorption of Ly$\alpha$ by the neutral intergalactic medium (IGM) at the highest redshifts \citep{Treu2013a}. Fortunately, studies in longer wavelength emission are not so severely impacted by the neutral IGM. 

IR emission lines like [CII] (158$\mu$m) serve as a robust redshift tracer of the interstellar medium. Recent studies of atomic gas in the vicinity of high-redshift quasars include \citet{Decarli2017a}, which discovered an overdensity of galaxies in [CII] around quasars in a sample that overlaps with that presented here. From this sample, Decarli et al. present bright, gas-rich, star-forming galaxies at the same redshift as the quasars in 4 fields (two of which also have strong continuum emission) concluding that this is consistent with a local peak in the overall galaxy number counts in [CII]. ALMA observations have revealed a number of millimeter continuum and [CII] companions near $z\sim5$ \citep{Trakhtenbrot2018a} and $z\sim6$ quasars \citep{Willott2017a}. Here, we focus on the number counts of dust continuum sources surrounding the quasar environment, which provide a direct measure of star formation activity in galaxies at $z\sim6$. Millimeter continuum allows us to probe a larger redshift range than in [CII], which can additionally be very faint compared to the FIR luminosity. Searching specifically for dust continuum sources should provide a complementary dataset to [CII] studies that could trace these detected overdensities, as it is likely that galaxies with strong [CII] signatures indicative of star formation would contain a significant amount of dust.

While [CII] provides precise redshift information, dust statistics probe a much larger line of sight volume, allowing for considerable foreground projection. However, the number counts for 1.1\,mm and 1.2\,mm sources are more certain than high-redshift UV, IR, and [CII] luminosity functions \citep[e.g.,][]{Finkelstein2015a, Aravena2016b}, so we would not need to detect an extreme (i.e., orders of magnitude) excess of galaxies in order to find robust evidence for protoclusters. Without spectroscopic redshifts, the hypothesis of this study is that a high number of millimeter dust emitters found in close angular proximity to the high-redshift quasars, as compared with randomly chosen blank fields, could provide statistical evidence for clustering at early times. Other studies lacking redshifts, including radio continuum \citep[e.g.][]{Carilli2004a} and using the Lyman-break technique \citep[e.g.][]{Kim2009a}, have been unable to confirm physical association with the quasar, but an overdensity of sources beyond small number statistics would be evidence for early protoclusters, similar to the LBG overdensity detected in the protocluster TN J1338 \citep{Miley2004a} as well as the submillimeter mapping of radio galaxies and their companions \citep{Stevens2003a}. 

%Continuum observations around radio-loud galaxies have yielded evidence for early clustering \citep[e.g.][]{Carilli2004a, Venemans2007b, Wang2008a, Wylezalek2013a}, but 
The increasing sample size of high redshift quasars opens a particularly interesting avenue for protocluster detections at longer wavelengths, in complement to the LAE searches. The number of confirmed bright $z>6$ quasars has significantly increased recently through large-area sky surveys \citep[e.g.][]{Banados2014a, Banados2016a, Jiang2016a, Matsuoka2016a, Mazzucchelli2017b, Reed2017a}, and many of these have been observed in millimeter wavelengths \citep[e.g.][]{Decarli2018a}. Thus, for the first time we have a sufficiently large statistical basis for a search for possible overdensities of dusty, gas-rich galaxies in their surrounding environment at very high redshifts. Additionally, the ALMA era has made it possible to find faint sources in continuum at very high redshift ($z>5-6$). 

In this study we examine the primary beams of high-redshift quasar pointings from ALMA Cycles 0$-$4 and compare them with the number counts of sources in blind surveys, providing a unique constraint on whether dust continuum traces the overdensities already known to be present in [CII]. In section 2 we describe the observations and data reduction; in section 3 we give details on the analysis of the images and calculation of the number counts; and in sections 4 and 5 we give a discussion and concluding remarks. We assume the \textit{Planck} cosmology with a flat Universe, with $\rm H_0$ = 67.77 $\rm km\, s^{-1} Mpc^{-1}$ and $\Omega_m$ = 0.308 \citep{Planck2016a}.

\section{Observations and Data Reduction}

\begin{figure}
    \includegraphics[width=0.99\columnwidth]{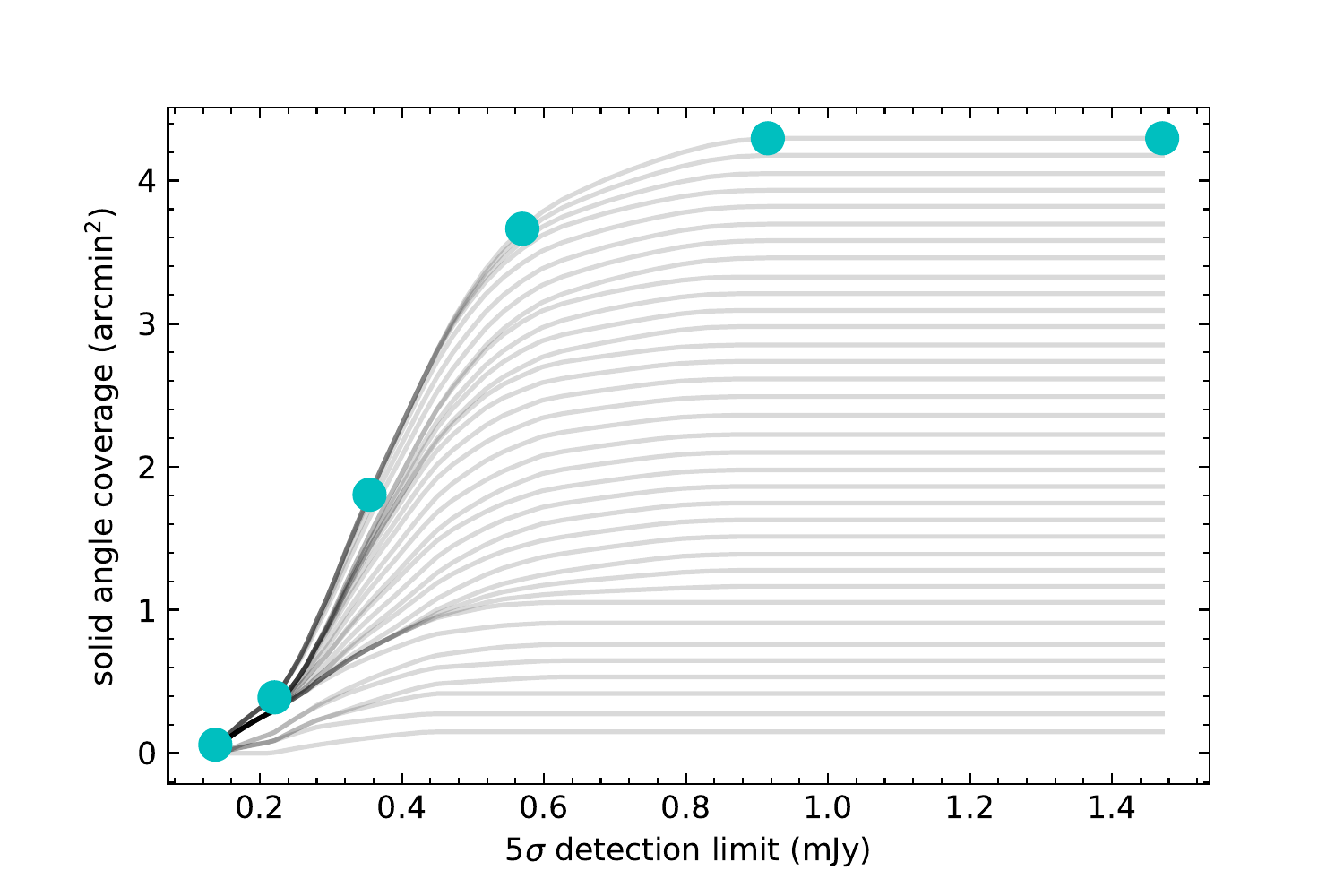}
    \caption{Area covered by this survey as a function of the flux to which it is sensitive at 5$\rm \sigma$. Each gray line represents the cumulative solid angle covered by each additional quasar, up to the full sample of 35. The points are binned according to the areas used in calculating the number counts.}
    \label{fig:area}
\end{figure}

\begin{deluxetable*}{lcccccccc}
\tablenum{1}
\tablecolumns{8}
\tablewidth{0pt}
\tablehead{
\colhead{Target name} & \colhead{Short name} & 
\colhead{RA} & \colhead{Dec} & \colhead{$\rm \sigma_{cont}$} & \colhead{$z$} & \colhead{Redshift method} & \colhead{Ref.}\\
& & \colhead{(J2000)} & \colhead{(J2000)} & \colhead{($\rm \mu$Jy)} & & & \\
\colhead{(1)} & \colhead{(2)} & \colhead{(3)} & \colhead{(4)} & \colhead{(5)} & \colhead{(6)} & \colhead{(7)} & \colhead{(8)}
}

\startdata
    
    PSO J007.0273+04.9571 &   PJ007+04 & 00:28:06.56 & $+$04:57:25.68 & 87.4 & 6.00$\pm$0.05 & [CII] & 1 %BANADOS16
    \\
    PSO J009.7355--10.4316 &   PJ009--10 & 00:38:56.522 & --10:25:53.90 & 84.9 & 5.95$\pm$0.05 & [CII] & 1 %BANADOS16
    \\
    VIK J0046--2837 & J0046--2837 & 00:46:23.65 & --28:37:47.34 & 70.8 & 5.9900$\pm$0.05 & [CII] & 2%VENEMANS--IP
    \\
    VIK J0109--3047 & J0109--3047 & 01:09:53.13 & --30:47:26.3 & 44.2 & 6.7909$\pm$0.0004 & [CII] & 15\\%VENEMANS17\\
    
    %SDSS J0129--0035 & 01:29:58.51 & --00:35:39.7 & 7 & 47.3 & 5.7794$\pm$0.0008 & CO & 15\\%WANG13\\
    
    ATLAS J025.6821--33.4627 & J0142--3327 & $+$01:42:43.73 & --33:27:45.47 & 61.2 & 6.31$\pm$0.03 & Ly$\alpha$ & 3\\%CARNALL15\\
    
    CFHQS J0210--0456 & J0210--0456 & 02:10:13.19 & --04:56:20.9 & 28.3 & 6.4323$\pm$0.0005 & [CII] & 14\\%WILLOTT13\\
    
    VIK J0305--3150 & J0305--3150 & 03:05:16.92 & --31:50:56.0 & 44.0 & 6.6145$\pm$0.0001 & [CII] & 15\\%VENEMANS17\\
    
    PSO J065.4085--26.9543 &   PJ065--26 & 04:21:38.052 & --26:57:15.60 & 60.9 & 6.14$\pm$0.05 & [CII] & 1 \\ %BANADOS16
    
    PSO J065.5041--19.4579 &   PJ065--19 & 04:22:00.994 & --19:27:28.68 & 46.6 & 6.12$\pm$0.05 & [CII] & 1 \\ %BANADOS16
    
    VDES J0454--4448 & J0454--4448 & 04:54:01.79 & --44:48:31.1 & 46.1 & 6.10$\pm$0.01 & [CII] & 4\\%REED17\\
    
    SDSS J0842+1218 & J0842+1218 & 08:42:29.429 & $+$12:18:50.50 & 43.0 & 6.069$\pm$0.002 & MgII & 5\\%DEROSA11\\
    
    SDSS J1030+0524 & J1030+0524 & 10:30:27.098 & $+$05:24:55.00 & 53.4 & 6.308$\pm$0.001 & MgII &  6\\%KURK17\\
    
    PSO J159.2257--02.5438 &   PJ159--02 & 10:36:54.191 & --02:32:37.94 & 60.7 & 6.38$\pm$0.05 & [CII] & 1 \\%BANADOS16
    
    %SDSS J1044--0125 & 10:44:33.04 & --01:25:02.2 & 7 & 66.1 & 5.7824$\pm$0.0007 & CO & 15\\%WANG13\\
    
    VIK J1048--0109 & J1048--0109 & 10:48:19.086 & --01:09:40.29 & 60.0 & 6.6610$\pm$0.005 & MgII & 2\\%VENEMANS--IP\
    
    PSO J167.6415--13.4960 & PJ167--13 & 11:10:33.976 & --13:29:45.60 & 43.7 & 6.508$\pm$0.001 & MgII & 7 \\%VENEMANS--IP
    
    ULAS J1148+0702 & J1148+0702 & 11:48:03.286 & $+$07:02:08.3 & 55.4 & 6.339$\pm$0.001 & MgII & 8\\%VENEMANS15\\
    
    VIK J1152+0055 & J1152+0055 & 11:52:21.269 & $+$00:55:36.69 & 54.5 & 6.3700$\pm$0.01 & Ly$\alpha$ & 9\\%MATsuoka16\\
    
    ULAS J1207+0630 & J1207+0630 & 12:07:37.440 & $+$06:30:10.37 & 53.1 & 6.040$\pm$0.003 & Ly$\alpha$& 8\\%Jiang15\\
    
    PSO J183.1124+05.0926 &   PJ183+05 & 12:12:26.981 & $+$05:05:33.49 & 62.3 & 6.4386$\pm$0.0004 & [CII] & 1\\ %BANADOS16
    
    SDSS J1306+0356 & J1306+0356 & 13:06:08.258 & $+$03:56:26.30 & 57.5 & 6.016$\pm$0.002 & MgII & 6\\%KURK07\\
    
    ULAS J1319+0950 & J1319+0950 & 13:19:11.29 & $+$09:50:51.4 & 60.7 & 6.133$\pm$0.0012 & CO &  13\\%KURK17\
    
    PSO J217.0891--16.0453 &   PJ217--16 & 14:28:21.394 & --16:02:43.29 & 68.9 & 6.11$\pm$0.05 & [CII] & 1 \\%BANADOS16
    
    CFHQS J1509--1749 & J1509--1749 & 15:09:41.778 & --17:49.26.80 & 57.9 & 6.121$\pm$0.002 & MgII & 10\\%WILLOTT10\\
    
    PSO J231.6576--20.8335 &   PJ231--20 & 15:26:37.841 & --20:50:00.66 & 98.4 & 6.5950$\pm$0.015 & [CII] & 11\\%MAZZUCHELLI17
    
    PSO J308.0416--21.2339 &   PJ308--21 & 20:32:09.996 & --21:14:02.31 & 27.7 & 6.24$\pm$0.05 & [CII] & 1 \\%BANADOS16
    
    SDSS J2054--0005 & J2054--0005 & 20:54:06.49 & --00:05:14.8 & 30.7 & 6.0391$\pm$0.0022 & CO & 13\\%WANG13\\
    
    CFHQS J2100--1715 & J2100--1715 & 21:00:54.616 & --17:15:22.50 & 49.0 & 6.087$\pm$0.005 & MgII & 10\\%WILLOTT10\\
    
    VIK J2211--3206 & J2211--3206 & 22:11:12.391 & --32:06:12.94 & 48.3 & 6.3360$\pm$0.005 & MgII & 2\\  %VENEMANS--IP\\
    
    PSO J340.2041--18.6621 &   PJ340--18 & 22:40:48.997 & --18:39:43.81 & 61.1 & 6.01$\pm$0.05&  [CII] & 1 \\%BANADOS16
    
    SDSS J2310+1855 & J2310+1855 & 23:10:38.88 & $+$18:55:19.7 & 51.7 &  6.0031$\pm$0.0007 & CO &  13\\%WANG13\\
    
    VIK J2318--3029 & J2318--3029 & 23:18:33.100 & --30:29:33.37 & 94.4 & 6.1200$\pm$0.05 & [CII] & 2\\%VENEMANS--IP
    
    VIK J2318--3113 & J2318--3113 & 23:18:18.351 & --31:13:46.35 & 86.7 & 6.4440$\pm$0.005 & MgII & 2\\%VENEMANS--IP
    
    CFHQS J2329--0301 & J2329--0301 & 23:29:08.28 & --03:01:58.8 & 21.0 & 6.417$\pm$0.002 & MgII & 14\\%WILLOTT13\\
    
    VIK J2348--3054 & J2348--3054 & 23:48:33.34 & --30:54:01.0 & 51.9 & 6.9018$\pm$0.0007 & [CII] & 15\\%VENEMANS13\
    
    PSO J359.1352--06.3831 &   PJ359--06 & 23:56:32.455 & --06:22:59.26 & 94.1 & 6.15$\pm$0.05 & [CII] & 1\\%BANADOS16
 
\enddata

%\ref{fig:tab1}
\tablecomments{(1) Original target name; (2) short name used throughout this paper; (3-4): RA \& Dec (J2000); (5) continuum RMS before primary beam correction ($\rm \mu$Jy); (6) redshift; (7) method for redshift determination, (8) redshift references.}

\tablerefs{1- \citet{Banados2016a};
2- Venemans et al. (in prep); 3- \citet{Carnall2015a}; 4-\citet{Reed2017a}; 5- \citet{De-Rosa2011a}; 6- \citet{Kurk2007a}; 7-
\citet{Venemans2015a}; 8- \citet{Jiang2016a}; 9- \citet{Matsuoka2016a}; 10- \citet{Willott2010a}; 11- \citet{Mazzucchelli2017a};
12- \citet{Kurk2009a}; %13- \citet{Becker2015a}; 
12- \citet{Willott2009a}; 
13- \citet{Wang2013a}; 14- \citet{Willott2013a}; 15- \citet{Venemans2013a}.}

\end{deluxetable*}
   
We examine 35 quasars spanning $5.95<z<6.90$ with ALMA observations in Band 6 (211-275 GHz, corresponding to rest-frame FIR $\sim150-200\mu$m). Eight of the quasars are archival observations from Cycles 0 and 1; the remaining 27 objects constitute the sample used in the [CII] search by \citet{Decarli2017a, Decarli2018a}. Thus, these observations have spectral windows encompassing the expected observed frequency of [CII]; they were carried out between 2016 January and July, with a compact configuration using 38-49 12\,m antennas. The primary beam of ALMA in Band 6 is about 25$\arcsec$ at FWHM, corresponding to an angular diameter distance of $\sim$140 physical kpc, or $\sim$1.0 comoving Mpc, at $z\sim6.5$, a representative redshift for the sample. 

The quasars in our sample were originally identified in multiple optical and near-infrared surveys including SDSS \citep{Jiang2016a}, Pan-STARRS1 \citep{Banados2016a}, VIKING \citep{Venemans2013a}, ATLAS \citep{Carnall2015a}, CFHQS \citep{Willott2007a}, UKIDSS \citep{Mortlock2009a}, and VDES \citep{Reed2017a}. The sample includes the 27 ALMA observations from \citet{Decarli2017a} as well as 8 quasars with archived observations from Cycles 0-1; this study includes every $z\sim6$ quasar with available Band 6 observations but is not formally complete in any physical quantity. These quasars were selected via detection in rest-frame ultraviolet at 1450$\rm \AA$ (absolute magnitudes $m_{1450\rm \AA} <-$25 mag), and many of them have had extensive sub/millimeter followup and most are hallmarked by strong [CII] detections and very bright ($S_{\nu} >$\,1 mJy) continuum emission in followup observations. Table 1 lists the information on the quasars and the details on the ALMA observations used here. 

The data have been reduced using the pipeline developed for the Common Astronomy Software Applications \citep[CASA;][]{McMullin2007a} and deconvolved using the CLEAN task. Each data set is imaged using natural weighting and is cleaned down to the 3$\rm \sigma$ noise level. The pixel size is $0\farcs1\times0\farcs1$ for the Cycle 4 data and $0\farcs15\times0\farcs15$ for the earlier cycles. The average RMS noise in the images is 58.6$\pm$19.0 $\mu$Jy\,beam$^{-1}$, spanning a range of 21.0$-$98.4~$\mu$Jy\,beam$^{-1}$; this is the uncorrected RMS calculated from the image after the quasar has been masked out but before we make the primary beam correction. The typical FWHM of the synthesized beam at these frequencies ranges from $0\farcs5$ to $1\farcs0$. We utilize the original resolution of each map, since homogenizing the resolution across the maps would cause an overall loss of depth. Moreover, the sizes of submillimeter galaxies are expected to be $\theta_{\rm FWHM} <0\farcs5$ at $z\sim2.5$ \citep{Hodge2016a}, therefore we expect all sources to be unresolved given these beam sizes.

We search for sources only where the primary beam response is greater than 50\% so the final images have been trimmed to this area (approximately $12\arcsec$ in radius). Figure \ref{fig:area} shows the total solid angle coverage of our search as a function of the 5$\sigma$ detection threshold, accounting for the primary beam and that each observation has a different depth. We discuss in section 3 the choice of the $5\sigma$ detection threshold. The deepest data have a flux limit of 0.15\,mJy\,beam$^{-1}$ and the maximum effective area is 4.3~arcmin$^2$.

\section{Data Analysis}\label{sec:3}
   \begin{figure}
    \includegraphics[width=0.99\columnwidth]{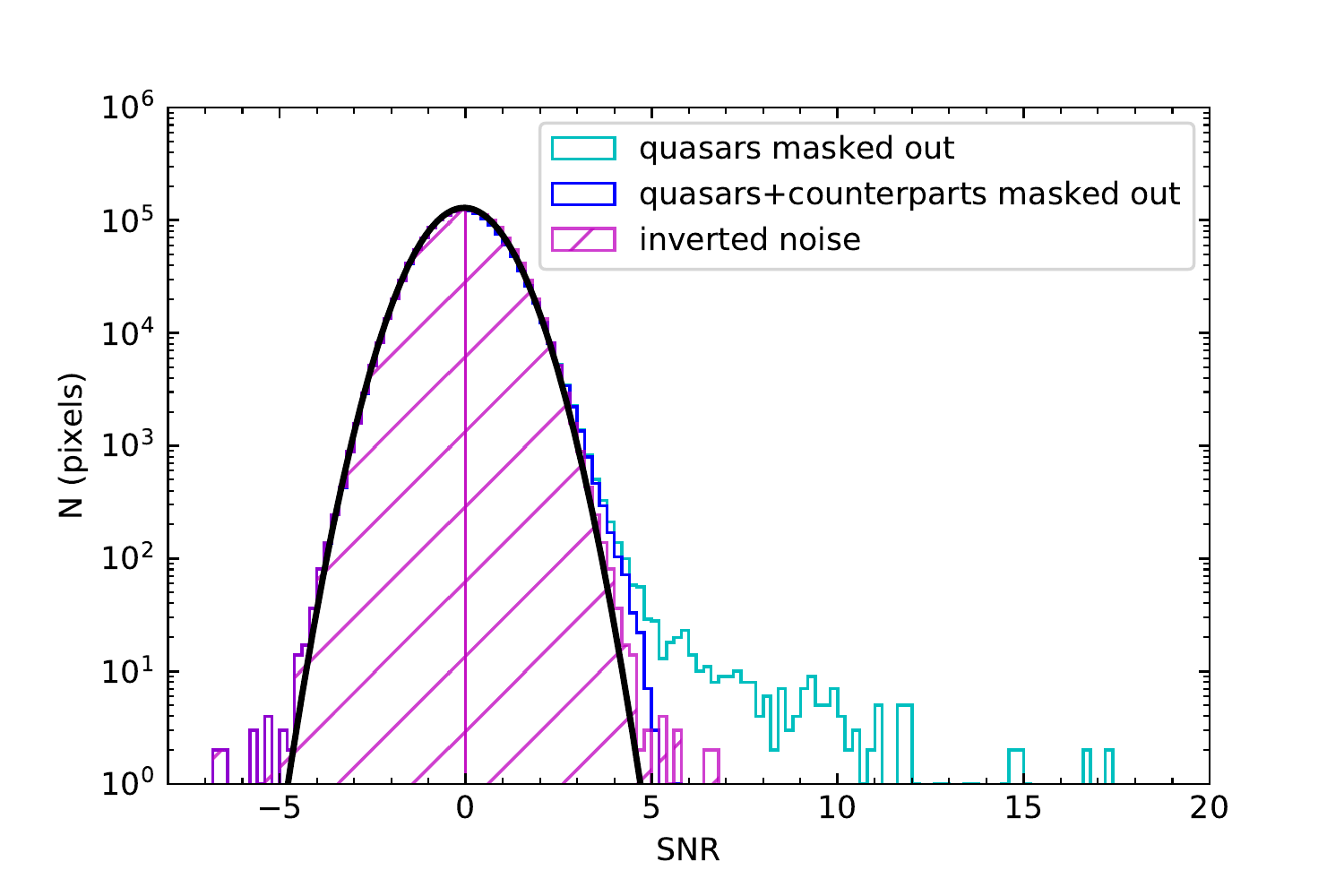}
    \caption{Distribution of the flux in the 35 fields after the target quasars have been masked out. The cyan histogram shows the total remaining flux in all of the quasar-masked maps; the blue histogram shows the remaining flux in the maps with both the quasar and the formally detected sources masked out (see section \ref{sec:3}); the hatched magenta histogram shows the negative flux inverted on the positive side to emphasize the positive excess in the total masked area. The black curve is a Gaussian fit to the noise with the expected $\mu=0$ and $\sigma=1$.}
    \label{fig:noisehist}
\end{figure}

\begin{figure}
    \includegraphics[width=0.99\columnwidth]{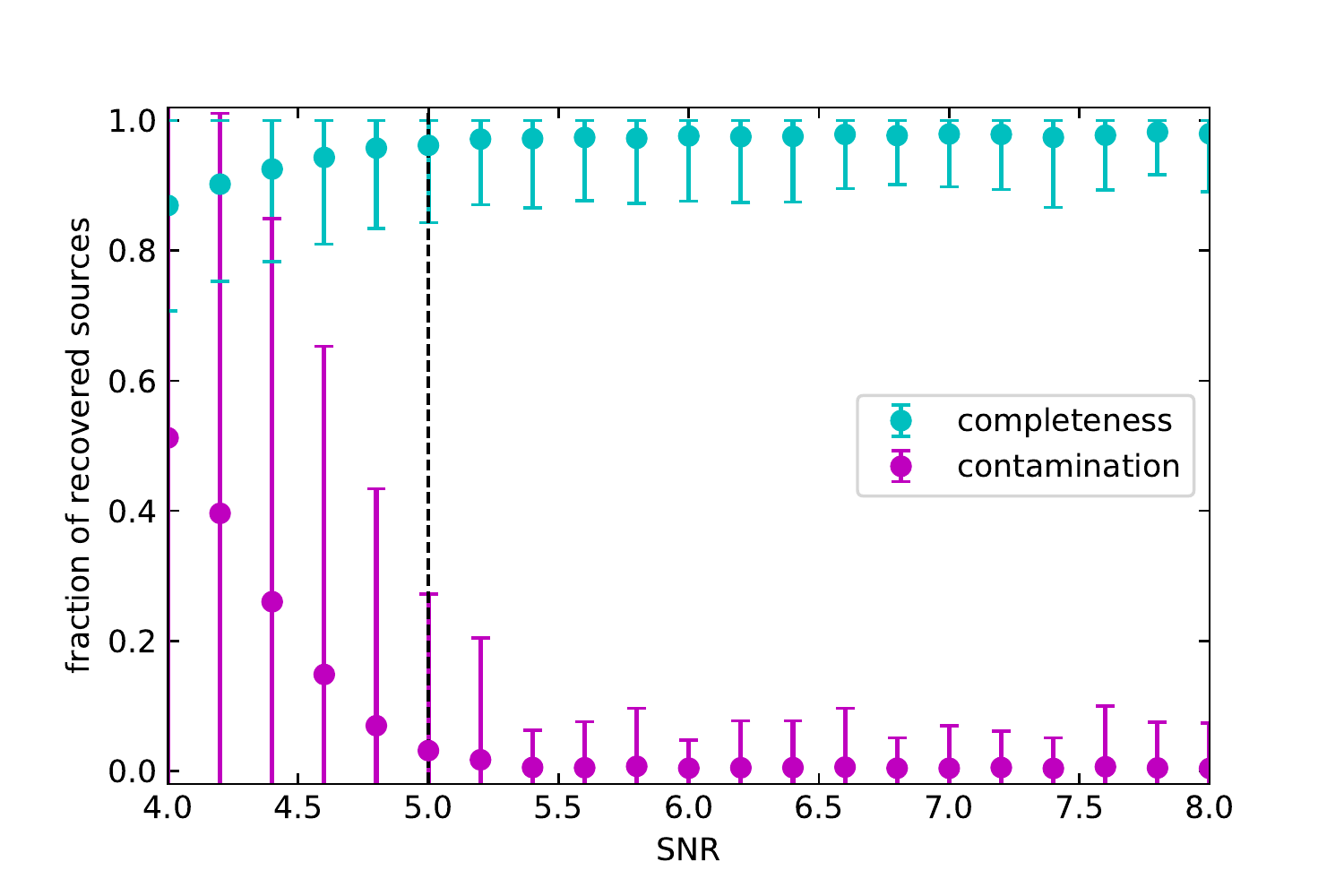}
    \caption{Results from 1000 simulated ALMA maps with sources injected according to blank field estimates. Magenta points show the contamination, i.e. the false detection rate. Cyan points indicate the ratio of recovered legitimate sources to the number of injected sources. The error bars are the 1$\rm \sigma$ standard deviation in the 1000 maps. We choose a SNR threshold of 5$\rm \sigma$ based on these results.}
    \label{fig:completeness}
\end{figure}

\begin{figure*}[ht!]
\begin{center}
    \includegraphics[width=1.0\columnwidth]{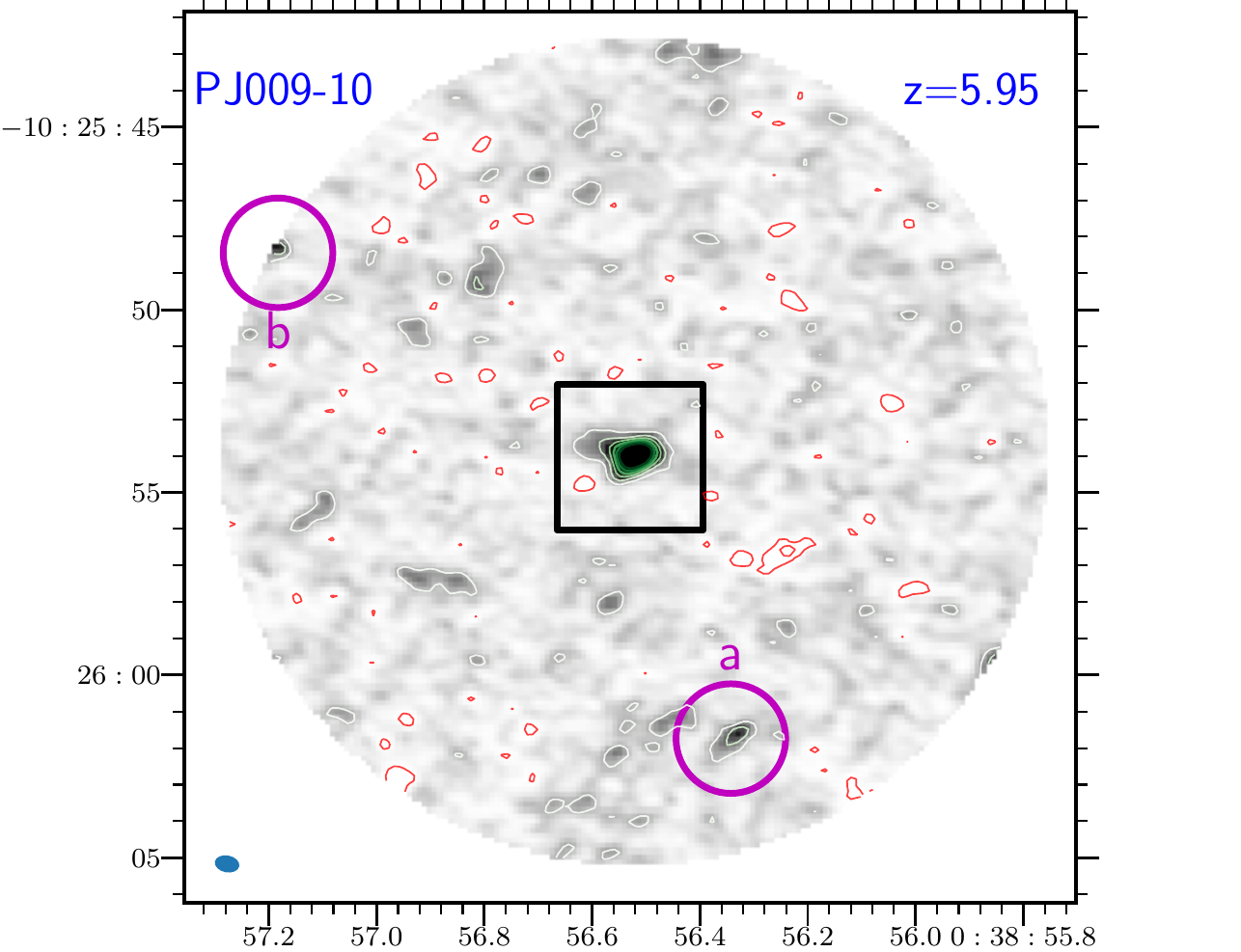}
    \includegraphics[width=1.0\columnwidth]{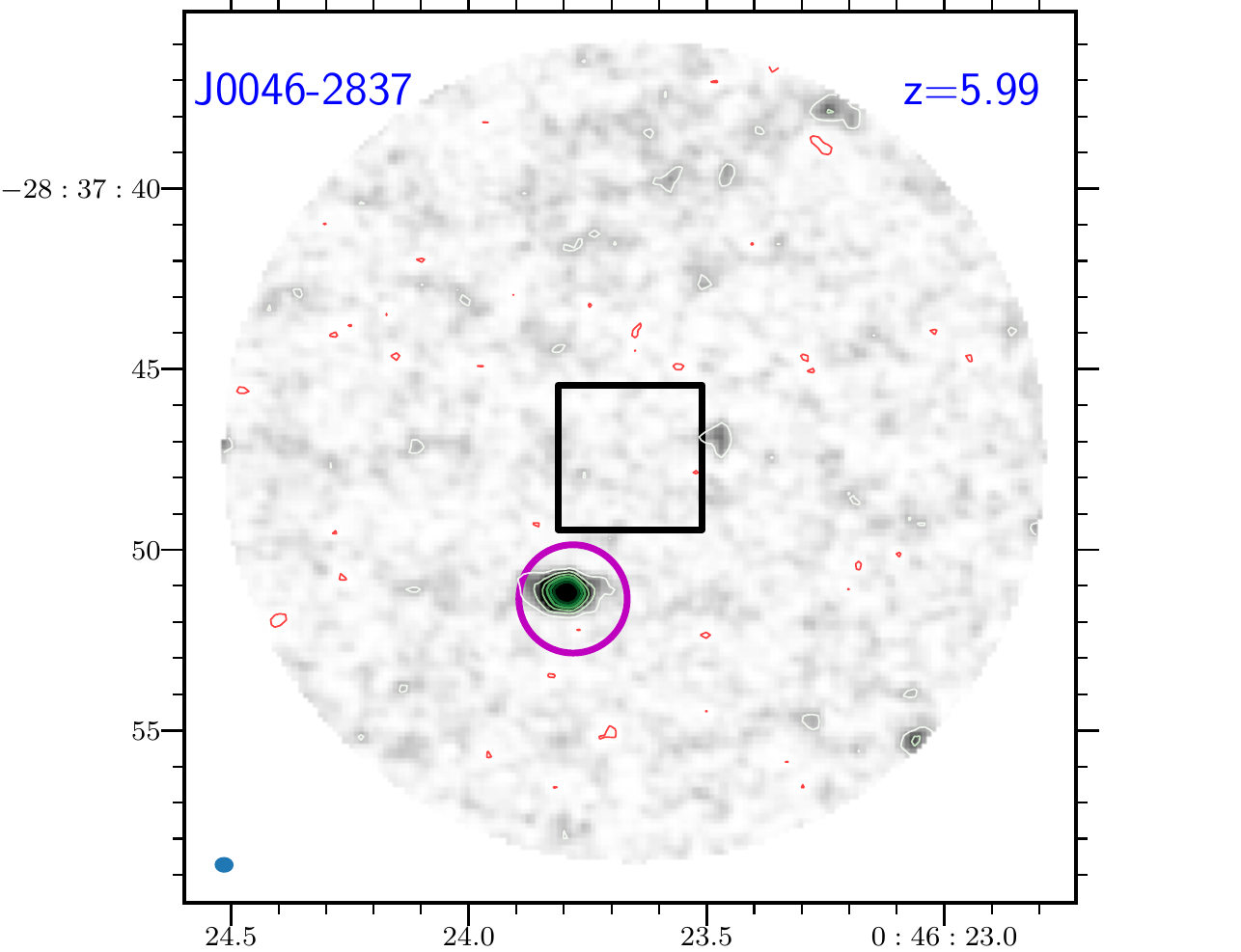}
    \includegraphics[width=1.0\columnwidth]{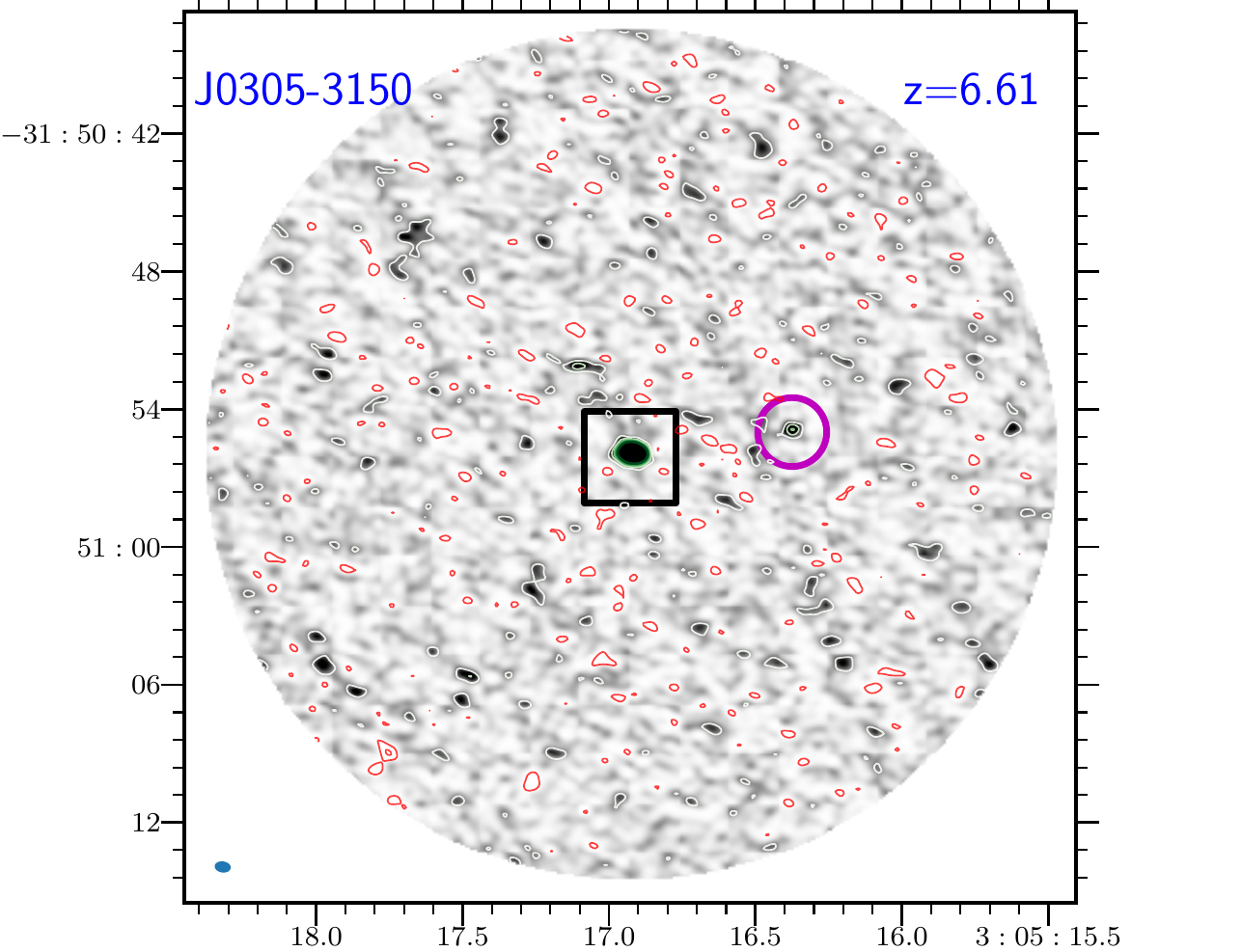}
    \includegraphics[width=1.0\columnwidth]{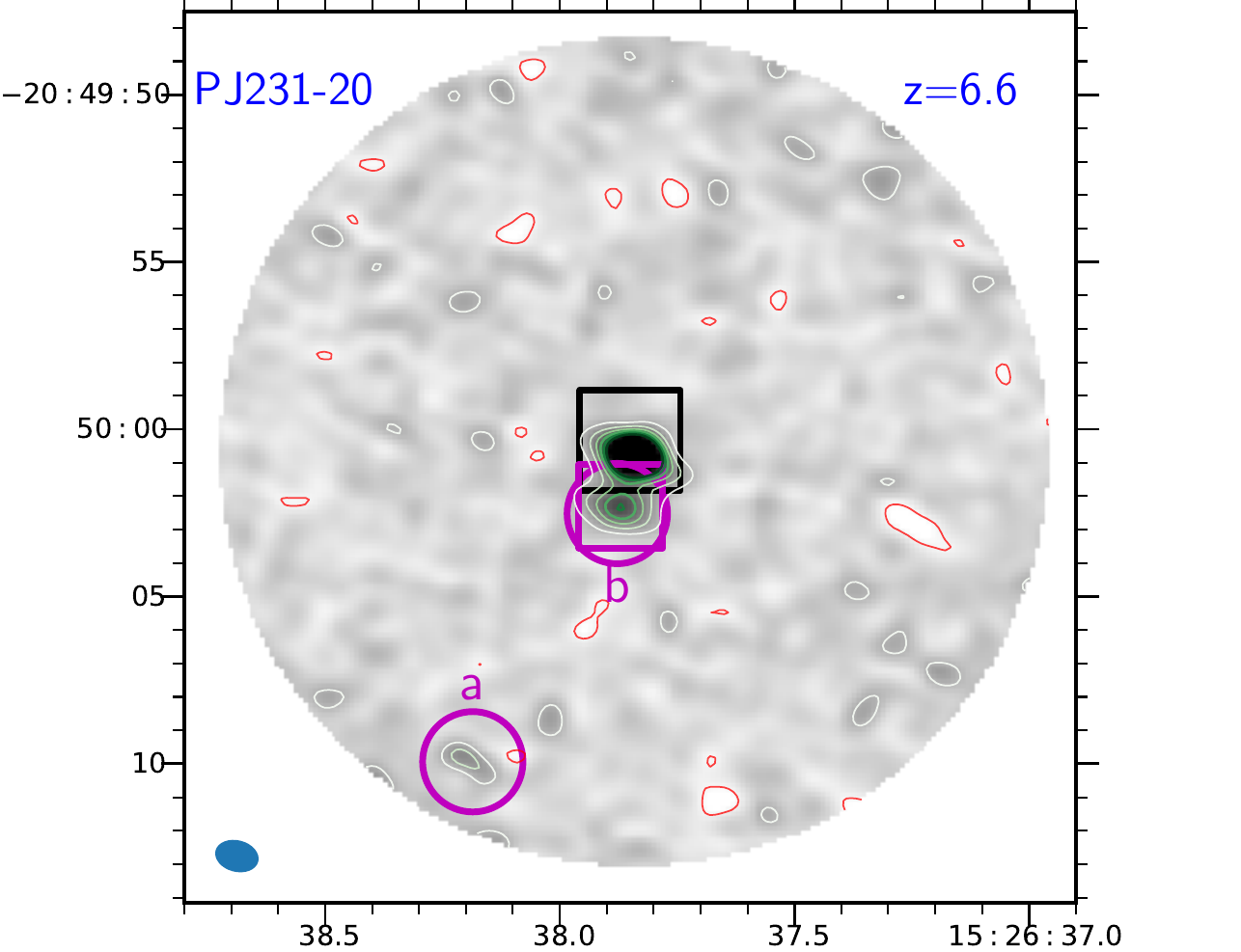}
    \includegraphics[width=1.0\columnwidth]{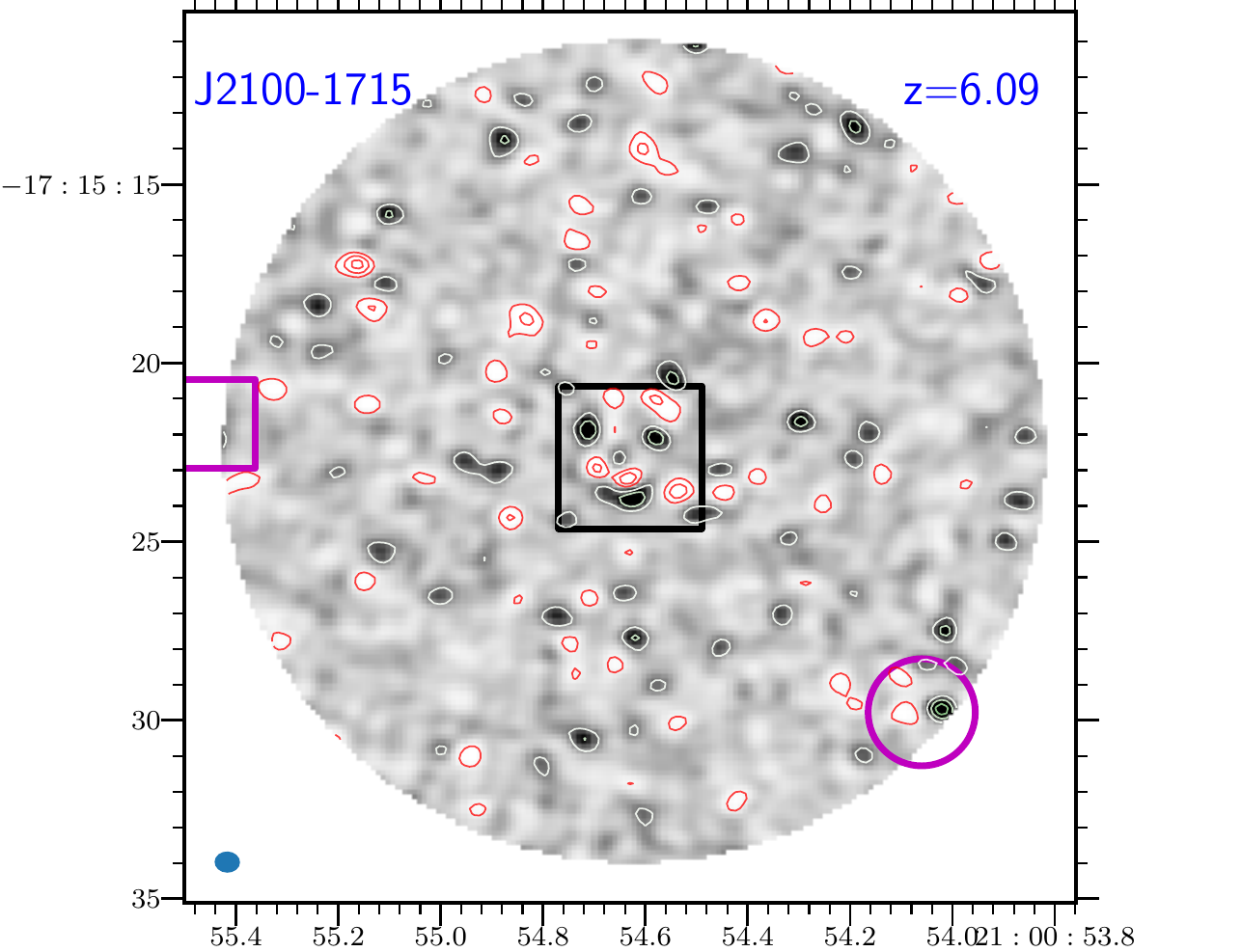}
    \includegraphics[width=1.0\columnwidth]{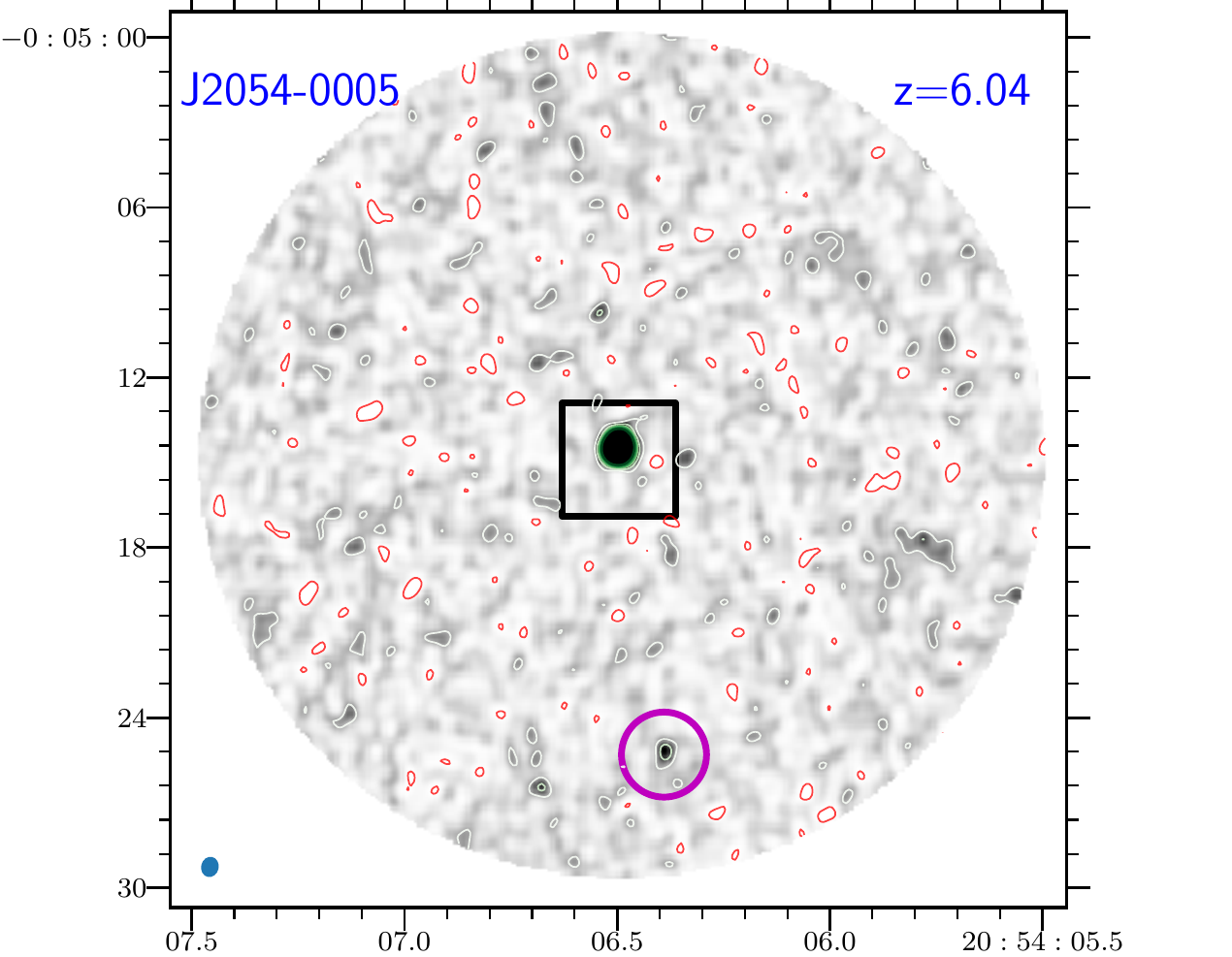}
    \caption{Each map is the cleaned ALMA continuum image within the area of 50\% of the primary beam response, with a 16\,arcsec$^2$ mask centered on the quasar. The synthesized beam is displayed in the lower left corner.
    The red (negative) and green (positive) contours begin at SNR=$\vert2\vert$ and step in increments of $\pm$2. 
    Magenta circles indicate $5\sigma$ continuum detections as listed in Table 2; squares show where \citet{Decarli2017a} found [CII] detections in the same fields. Note that the [CII] detection in the field of J2100-1715 corresponds to a continuum detection in our study but sits outside the 50\% primary beam cut, so is not included as a detection in Table 2. (cont. on next page)}
    \label{fig:cont1}
\end{center}
\end{figure*}

\begin{figure*}[ht!]
\begin{center}
    \includegraphics[width=1.0\columnwidth]{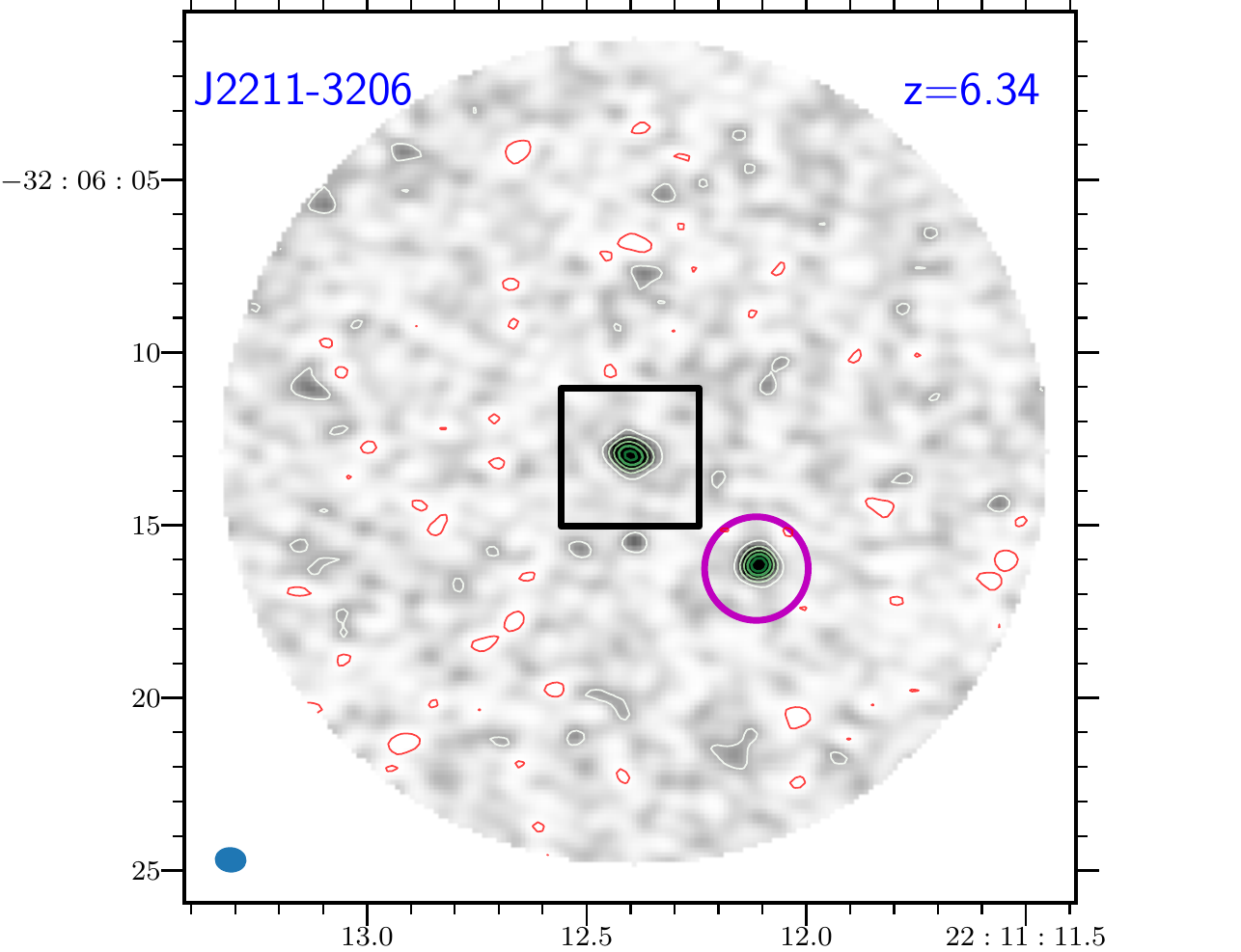}
    \includegraphics[width=1.0\columnwidth]{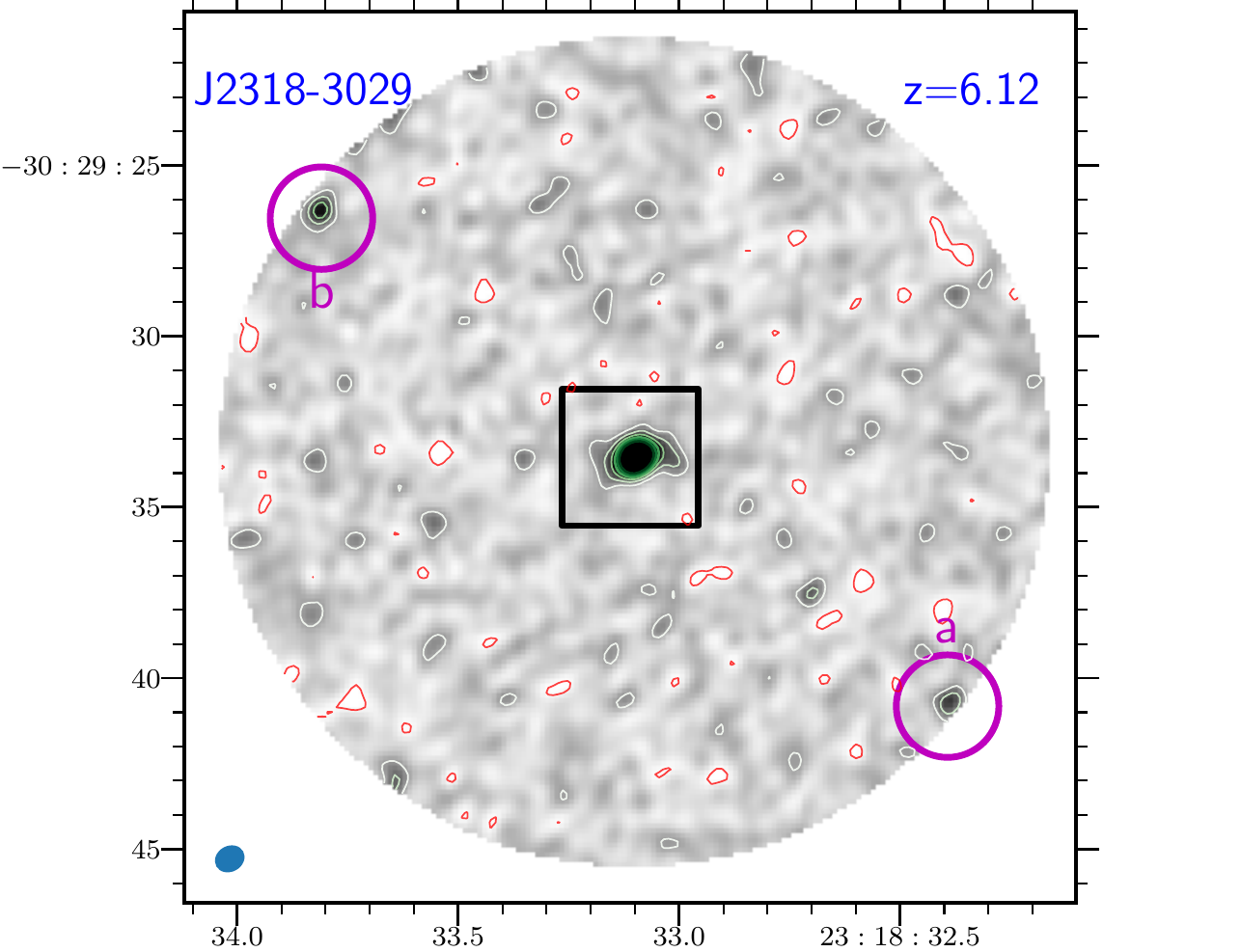}
    \includegraphics[width=1.0\columnwidth]{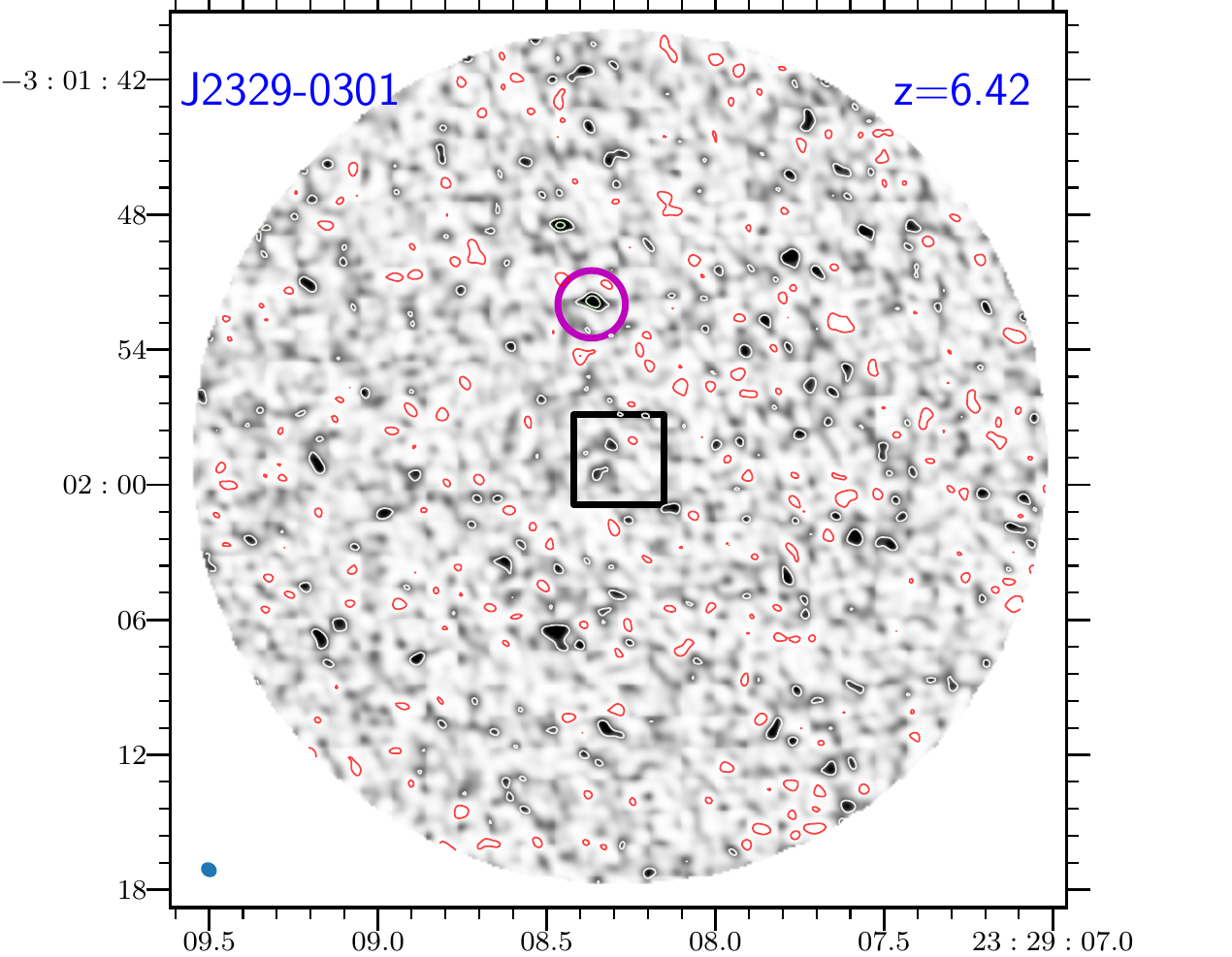}
    \includegraphics[width=1.0\columnwidth]{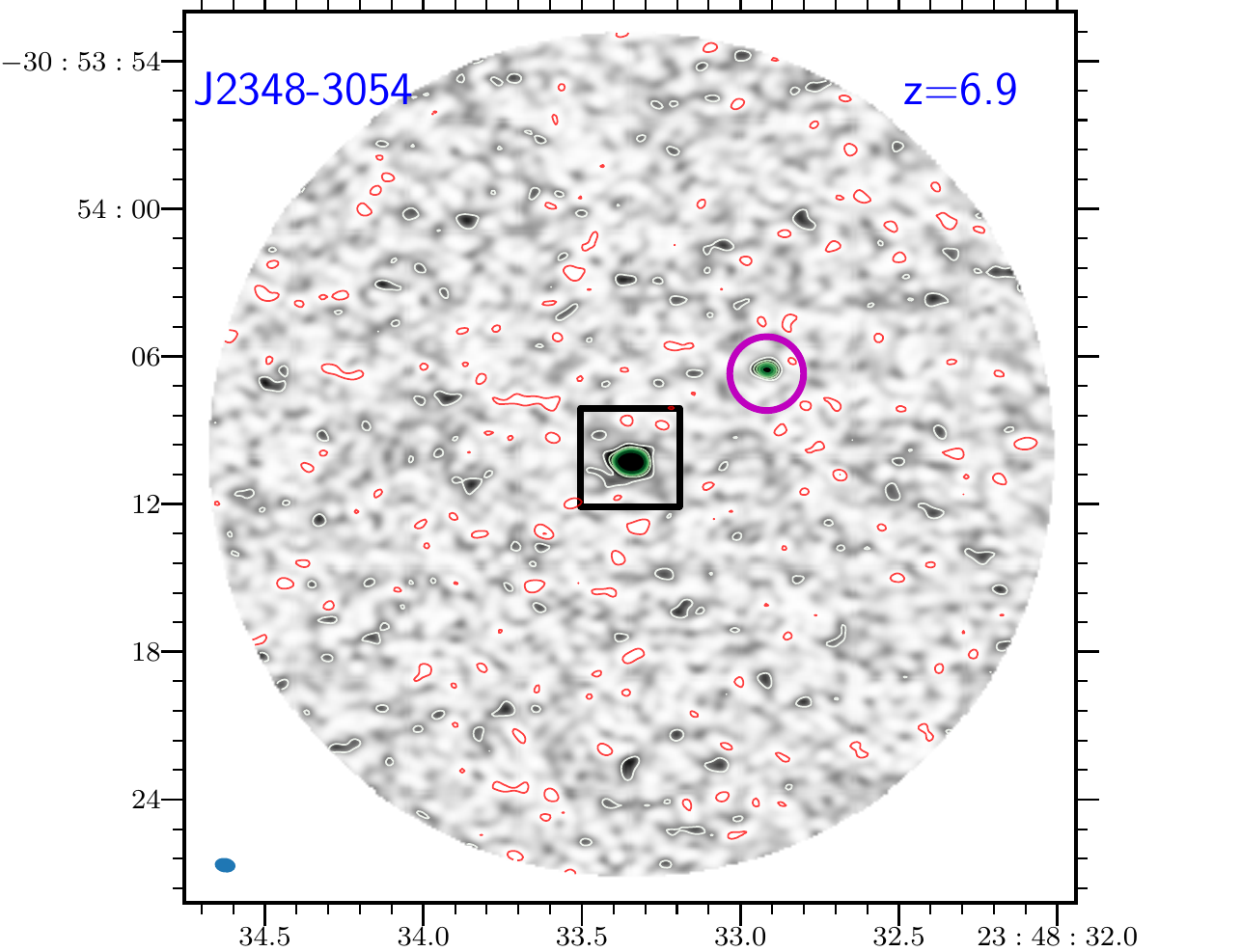}
    \includegraphics[width=1.0\columnwidth]{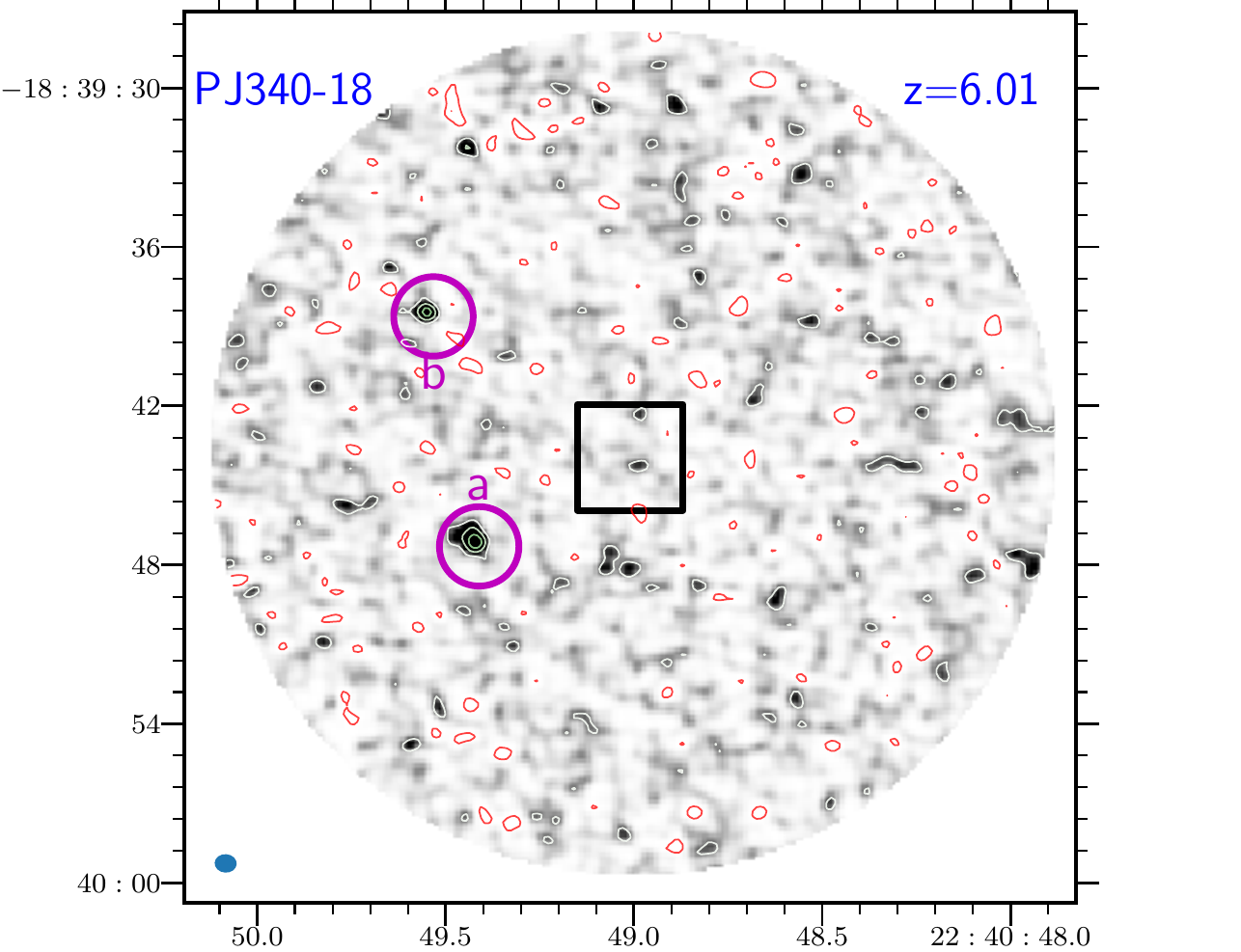}\\
    Figure \ref{fig:cont1} continued.
    %\caption*{Figure \ref{fig:cont1} continued.}
    %\label{fig:cont2}
\end{center}
\end{figure*}

We first search for a statistical excess of flux in each quasar field. We exclude the region of the field containing most of the quasar's flux by applying a conservative $4\arcsec \times 4\arcsec$ mask centered on the known quasar coordinates. In the case of J1152+0055, we increased the mask to a radius of 3$\arcsec$ to exclude extended emission from the quasar host galaxy, as our source-finding algorithm detected significant flux on the border of this mask; in the case of PJ231--20, there was a bright [CII] detection in \citet{Decarli2017a} within 2$\arcsec$ of the quasar so we decreased the mask to a radius of 1$\farcs$5. Besides PJ231--20, we consider as separate sources only detections at an angular distance $>2\arcsec$ ($\approx 14$\,pkpc) from the published position of the quasar. We perform this analysis on maps that have not been corrected for the primary beam, so that our source-finding algorithm does not interpret outer-edge pixels with higher noise to be true sources. This cut, in combination with a high signal-to-noise threshold, should prevent spurious noise detections; in addition, we address completeness and contamination via mock observations.

If there are zero sources in addition to the quasar in a cleaned primary beam, one would expect a Gaussian distribution of pure noise reflective of the RMS of the observations. To make a statistical statement about the excess flux in these fields, we construct a composite histogram of the total flux in the unmasked pixels in each of the maps. We convert the unmasked pixel fluxes to signal to noise (SNR) to account for both the primary beam and the different depths of each observation. In Figure \ref{fig:noisehist} we show the distribution of the flux in these masked maps: since negative flux is guaranteed to be noise, we invert this distribution to show what a pure noise distribution would look like, as compared with the clear statistical positive excess of millimeter flux in all of the fields, with the tail extending fairly bright (SNR$>$5). We then search for discrete sources, since interferometic observations filter out diffuse emission through the synthesized beam. Figure \ref{fig:noisehist} also shows the distribution of the flux with both the quasar and the formally detected 5$\sigma$ sources masked out (this threshold is explained below); this distribution is consistent with a slight positive tail indicating a few sources below the 5$\sigma$ detection threshold, but we do not include them in our measurements due to our estimation of the contamination. The distribution in Figure \ref{fig:noisehist} suggests that our argument for searching for formally detected discrete sources rather than statistical flux excess is sound.

We create signal-to-noise ratio maps to identify potential candidates in the field by dividing by the continuum RMS (not corrected for the primary beam). In order to find an optimal SNR threshold, we generate 1000 simulated maps with area equivalent to the 35 primary beams in this study. These maps are created with a Gaussian noise distribution, with $\sigma$ allowed to vary within the range of the rms values in our data. They are convolved with a model ALMA beam of $0\farcs8"\times0\farcs5$ -- for simplicity, we use a single beam shape typical of these data rather than the full range, since recovering point sources was not sensitive to the beam size we chose. The maps are injected randomly with sources according to a \citet{Schechter1976a} function distribution representative of blank field number counts (it is not necessary to account for confusion since the rms in the maps is well above the ALMA confusion limit). We assume the injected sources are unresolved and unclustered. The Schechter model we choose is based on the most recent blank field 1.2\,mm number counts compilation of \citet{Fujimoto2016a} valid for flux densities of 0.1-1.5\,mJy:
\begin{equation}
\small{\rm \frac{dN}{dlogS} = \rm ln(10)\phi_*\Big{(}\frac{S}{S_*}\Big{)}^{\alpha + 1}exp\Big{(}-\frac{S}{S_*}}\Big{)}
\end{equation}
%\begin{equation}
    %\small{\frac{dN}{dS} = \frac{N_{3 mJy}}{S'}\Big{(}\frac{S}{S_0}\Big{)}^{\alpha+1} \rm exp{\frac{-(S-S_0)}{S'}}}   
%\end{equation}
%where $\rm N_{3 mJy}$ = 230 deg{$\rm ^{-2}$}, S' = 1.7 mJy, $\rm S_0$ = 3 mJy and $\alpha$ = -2. 
where $\phi_* = 1.54^{+0.29}_{-0.26}$\,$\times$10$^3$\,deg$^{-2}$\,mJy$^{-1}$, $\alpha=-2.12^{+0.07}_{-0.06}$, and $\rm S_* = 2.35^{+0.16}_{-0.16}$\,mJy. Figure \ref{fig:completeness} shows the completeness (fraction of recovered sources that match the injected sources) and the
contamination (fraction of extracted sources that are false) after searching the simulated noise maps for the injected sources using a region-growing algorithm. We convert each simulation from flux density to SNR by dividing by the RMS noise in each iteration, which is allowed to vary within 1$\rm \sigma$ from the average RMS in our real maps. The error bars reflect the 1$\rm \sigma$ standard deviation of the completeness or contamination in all 1000 maps. Figure \ref{fig:completeness} indicates that we reach a satisfactory completeness of 91.6\% and contamination of 4.3\% at a $5\sigma$ detection threshold. After choosing this SNR threshold, we feed the real SNR maps to the same region-growing algorithm which identifies neighboring pixels in the map above the given threshold. A grouping is declared a source if there are two or more adjacent pixels above the SNR threshold.

\section{Results}
The results from the source-finding algorithm reveal a total of 15 candidate detections in the 35 fields, none of which are resolved. Contour plots showing the source detections in the quasar-masked fields are displayed in Figure \ref{fig:cont1}, while three [CII] sources found in \citet{Decarli2017a} and \citet{Willott2017a} which are not found in continuum here are shown in Figure \ref{fig:cont3}; the rest of the maps which have no additional sources are displayed in Appendix A.  In Table 2 we list the coordinates and primary beam-corrected fluxes of each source. Since these are continuum flux densities, we lack redshifts for these sources so we do not make a statement about the total IR-luminosity or the star formation activity in these sources.

For comparison with what is observed in blank fields, we construct the number counts as in previous ALMA surveys \citep[e.g.][]{Fujimoto2016a}. The contribution of an identified source of flux density $S$ to the total number counts, $\xi$, is:

\begin{equation}
    \xi(S) = \frac{1 - f(S)}{C(S) A_{\rm eff}(S)}
\end{equation}

\noindent where $f(S)$ is the contamination factor at $5\sigma$ from Figure \ref{fig:completeness}, $C(S)$ is the level of completeness (Figure \ref{fig:completeness}), and $A_{\rm eff}$ is the effective area of our search at flux density $S$, from Figure \ref{fig:area}. 

The sum of the contributions in each bin of flux width $\Delta$log$S$ is thus:

\begin{equation}
    n(S) = \frac{\Sigma \xi(S)}{\Delta \rm logS}
\end{equation}

The number counts are thus constructed in 6 log bins of $\Delta$log$S_{\nu}=0.25$, divided by the area sensitive to that flux level and corrected for both completeness and contamination.  Table 3 lists the values used for the number counts in this work. The uncertainty is calculated based on Poisson statistics propagated with the 1$\sigma$ uncertainty in the completeness and contamination as from our simulations. We compare the distribution of our sources in Figure \ref{fig:nc} with several blank field number counts at 1.1 and 1.2\,mm, with fits from \citet{Scott2010a} and \citet{Fujimoto2016a}. The \citet{Fujimoto2016a} curve is given along with 16th and 84th percentile values calculated from a Monte Carlo simulation, and the results from our simulations are plotted with the same limits. 
We calculate an overdensity factor defining $\delta_{\rm gal} = (N_{\rm gal}-N_{\rm exp})/N_{\rm exp}$. According to the \citet{Fujimoto2016a} model, we would expect $N_{\rm exp} = 16.1^{+3.3}_{-6.8}$ sources between 0.1-10\,mJy in these fields without an overdensity, based on the effective area in each bin. Correcting for fidelity and completeness, our measured number of sources $N_{\rm gal}$ is 15.1$\pm$5.7.  Therefore, this implies a value of $\delta_{\rm gal} = -0.07\pm0.56$, consistent with the expected number of sources in the field. Thus, we find that 15 sources in a total area of 4.3 $\rm arcmin^2$ does not point to an overdensity of 1.2\,mm sources with respect to the foreground.

Figure \ref{fig:radial} shows the observed and expected number of sources as a function of radial distance from the quasar in a given field. $N_{\rm gal}$ has been recalculated with each field cut to some radius $r$ from the quasar (and the area corrected for sensitivity as before). The sources are consistent with a uniform distribution at 50\% of the primary beam, and this remains true at any area cut: the observed number of sources is consistent with the scatter of the expected number of observed sources regardless of area cut, thus showing that our choice of area has not influenced the calculation of the overdensity, and that our measurements are not the result of a poor cleaning procedure.

\begin{figure}[ht]
\begin{center}
    \includegraphics[width=0.99\columnwidth]{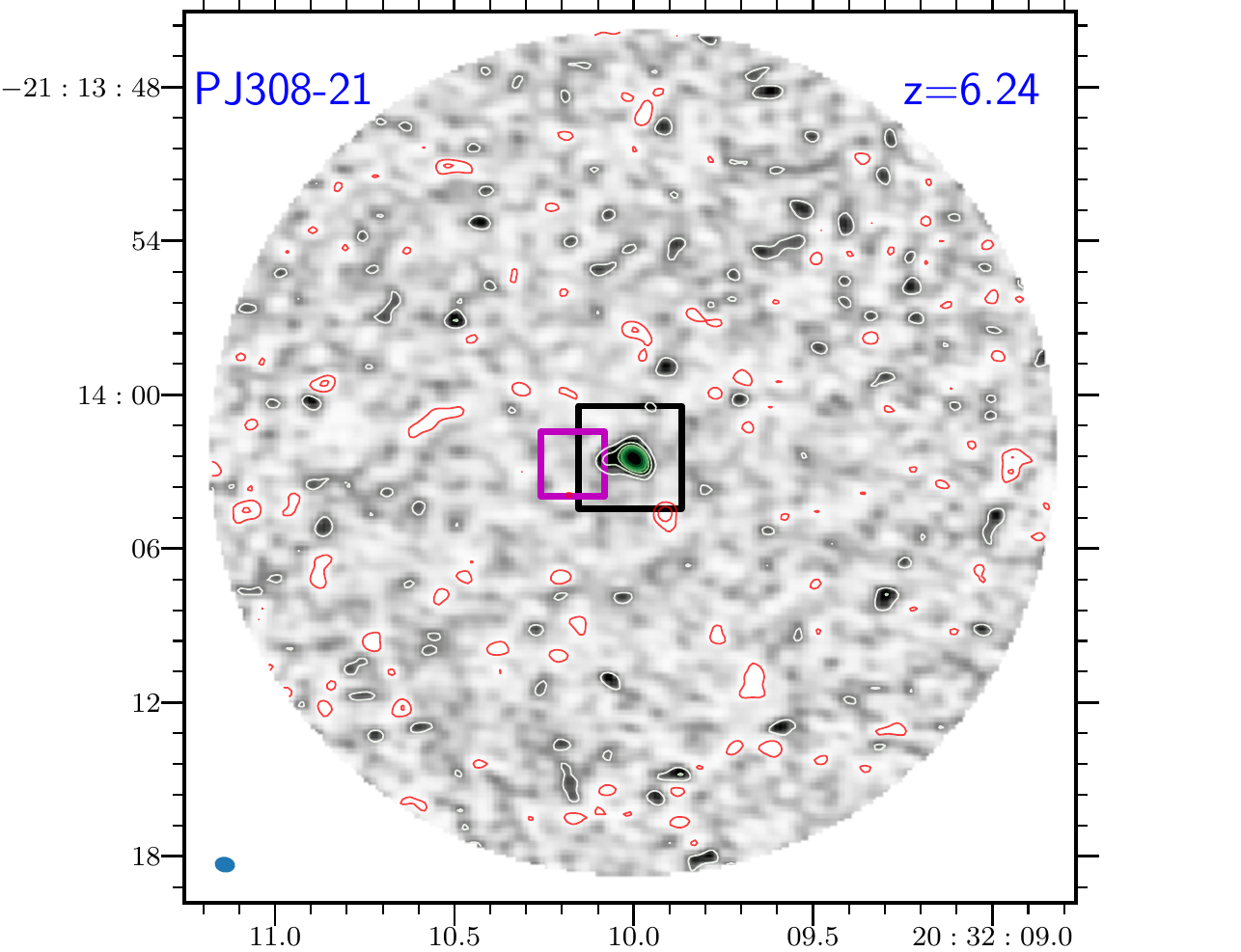}
    \includegraphics[width=0.99\columnwidth]{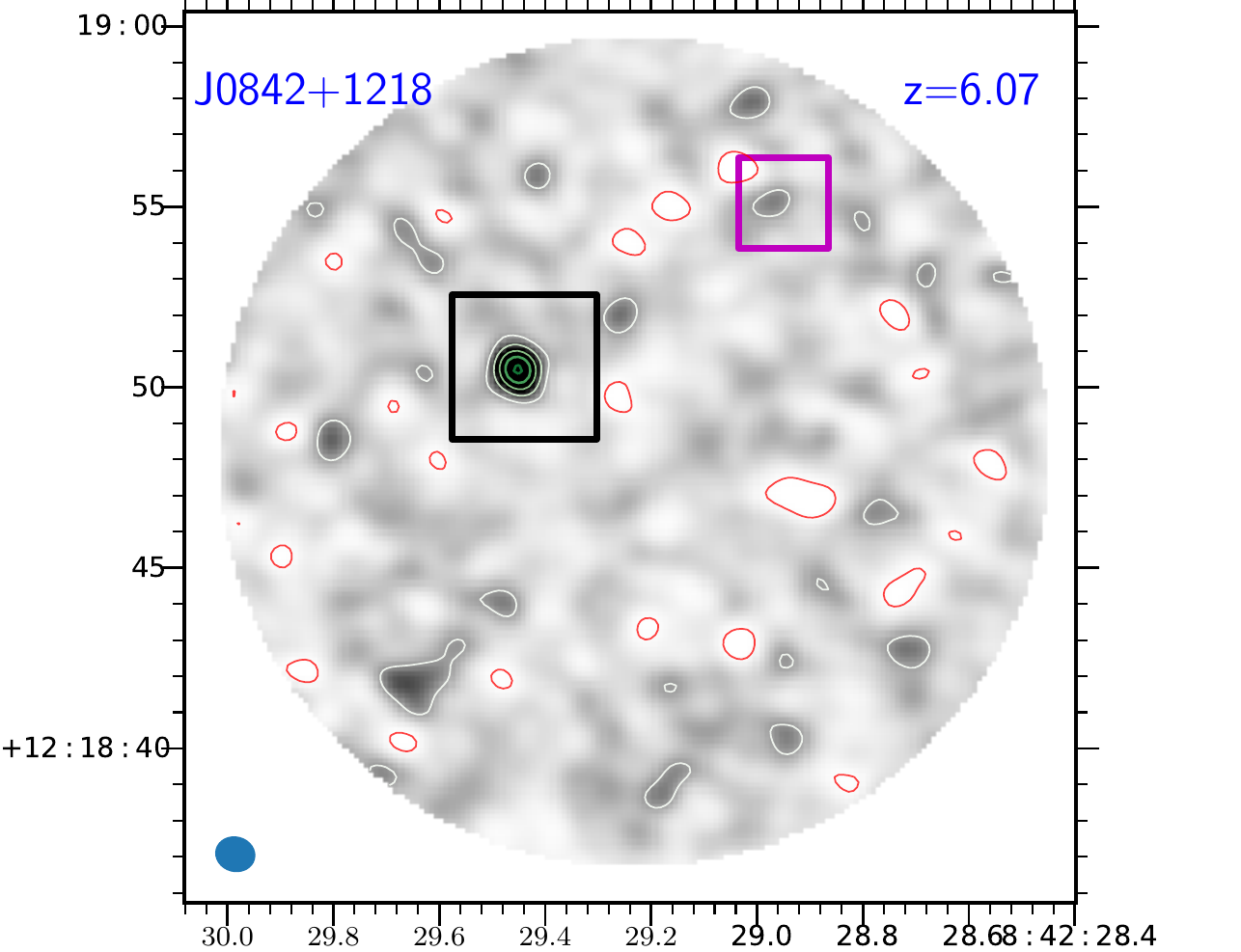}
    \includegraphics[width=0.99\columnwidth]{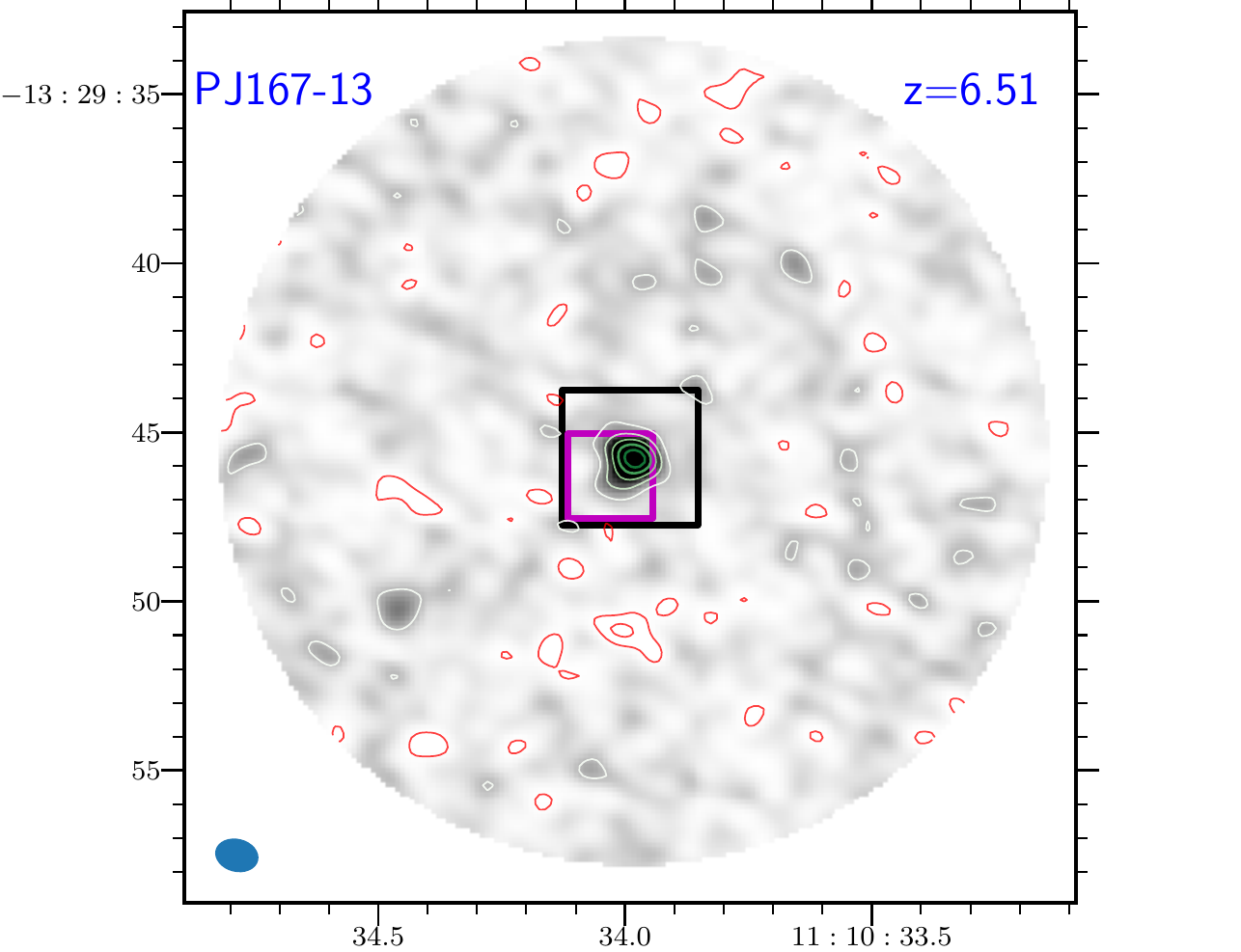}
    \caption{Three fields from this sample in which [CII] (magenta squares) was discovered but no $5\sigma$ dust continuum counterpart was discovered in this study. QSOs PJ308--21 and J0842+1218 mark detections from \citet{Decarli2017a}. QSO PJ167--13 marks a detection from \citet{Willott2017a}; the quasar has not been masked out in this map due to the close angular proximity of the companion source.}
    \label{fig:cont3}
\end{center}
\end{figure}

\begin{figure*}[ht]
\begin{center}
    \includegraphics{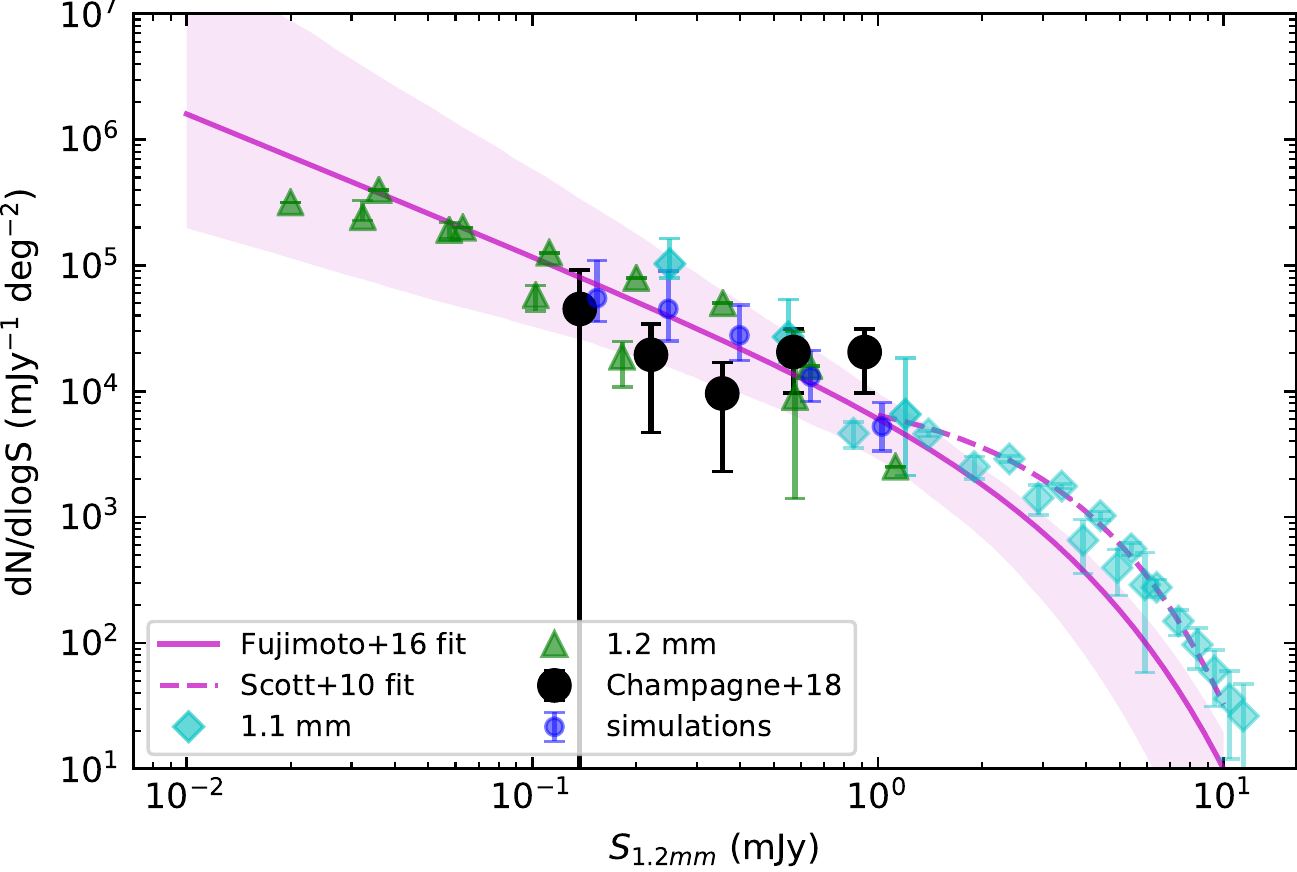}
    \caption{Blank field number counts from previous studies are shown as represented in the legend: \citet{ Aravena2016a, Fujimoto2016a, Hatsukade2016a, Oteo2016a, Scott2010a, Scott2012a}; the Schechter fits use the parameters from \citet{Fujimoto2016a}. The shaded magenta region represents the 16th and 84th percentile bounds on the MC simulation based on the \citet{Fujimoto2016a} function. The green points correspond to 1.2\,mm number counts; the cyan points correspond to 1.1\,mm number counts; the dark blue points correspond to an average over the 1000 simulated maps we have created according to \citet{Fujimoto2016a}; the black points are from this study. These points represent the total number of candidate companions divided by the area of the study to which that flux bin is sensitive. The black points are calculated in six log(flux) bins from 0.1-1.5\,mJy.}
    \label{fig:nc}
\end{center}
\end{figure*}

\begin{figure}
\begin{center}
    \includegraphics[width=1.0\columnwidth]{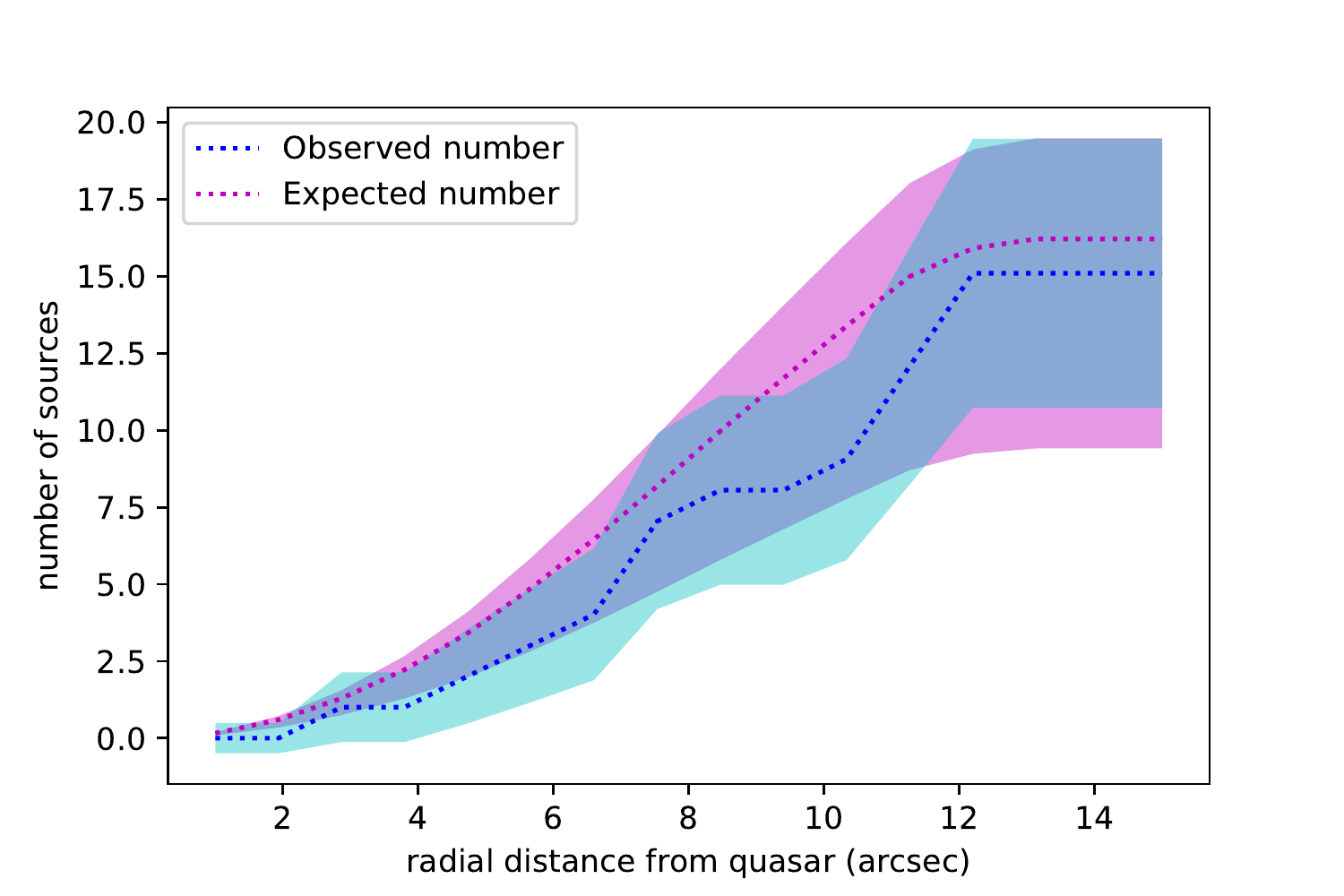}
    \caption{Number of sources as a function of radial distance from the quasar. $\rm A_{eff}$ has been corrected for sensitivity as before, but is calculated based on angular radius rather than fraction of the primary beam. The magenta line and shaded region shows the number of sources expected and its error based on our simulation; the cyan line and shaded region shows the observed number of sources and associated error. No overdensity arises at any angular separation, showing that our choice of primary beam cut has not influenced our calculation.}
    \label{fig:radial}
    \end{center}
    
\end{figure}

%\clearpage

\section{Discussion}

\begin{deluxetable*}{lccccccc}
\tablenum{2}
\tablecolumns{7}
\tablewidth{0pt}
\tablehead{
\colhead{Name} & \colhead{RA} & \colhead{Dec} & \colhead{Peak flux} & \colhead{SNR} & \colhead{PB resp.} & \colhead{Offset} \\
& \colhead{(J2000)} & \colhead{(J2000)} & \colhead{(mJy)} & & & \colhead{from Quasar ($\arcsec$)} \\
\colhead{(1)} & \colhead{(2)} & \colhead{(3)} & \colhead{(4)} & \colhead{(5)} & \colhead{(6)} & \colhead{(7)}
}

\startdata

C--PJ009a & 00:38:56.342 & --10:26:01.740 & 0.67$\pm$0.16 & 5.83 & 0.705 & 8.3 \\
C--PJ009b & 00:38:57.182 & --10:25:48.440 & 0.95$\pm$0.32 & 5.94 & 0.501 & 11.3 \\
C--J0046 & 00:46:23.780 & --28:37:51.360 & 1.47$\pm$0.09 & 17.36 & 0.916 & 4.5 \\
C--J0305 & 03:05:16.374 & --31:50:55.000 & 0.30$\pm$0.05 & 6.18 & 0.915 & 7.0 \\
C--PJ231a & 15:26:37.877 & --20:50:02.520 & 0.93$\pm$0.09 & 10.18 & 0.995 & 1.9 \\
C--PJ231b & 15:26:38.185 & --20:50:09.940 & 0.75$\pm$0.24 & 5.10 & 0.620 & 10.6 \\
C--J2054 & 20:54:06.390 & --0:05:25.299 & 0.23$\pm$0.06 & 5.22 & 0.716 & 10.6 \\
C--J2100 & 21:00:54.060 & --17:15:29.780 & 0.68$\pm$0.17 & 7.40 & 0.525 & 11.1 \\
C--J2211 & 22:11:12.113 & --32:06:16.260 & 0.64$\pm$0.06 & 11.74 & 0.903 & 4.8 \\
C--PJ340a$^{\dagger}$ & 22:40:49.410 & --18:39:47.300 & 0.51$\pm$0.08 & 7.48 & 0.891 & 7.1 \\
C--PJ340b$^{\dagger}$ & 22:40:49.531 & --18:39:38.620 & 0.53$\pm$0.10 & 6.86 & 0.795 & 9.6 \\
C--J2318a & 23:18:32.392 & --30:29:40.819 & 1.02$\pm$0.33 & 5.89 & 0.521 & 11.8 \\
C--J2318b & 23:18:33.809 & --30:29:26.539 & 1.23$\pm$0.32 & 7.26 & 0.532 & 11.6 \\
C--J2329 & 23:29:08.366 & --03:01:51.989 & 0.14$\pm$0.02 & 5.97 & 0.921 & 6.9 \\
C--J2348 & 23:48:32.917 & --30:54:06.700 & 0.69$\pm$0.06 & 11.96 & 0.919 & 6.5 \\
\enddata

    \tablecomments{Field sources identified in the thirty-five ALMA maps. (1) Name corresponds to the short names of the field in which they are detected. (2-3) RA, dec of each field source. (4) Peak flux is calculated from the brightest pixel in the candidate and is corrected for the primary beam offset. (5) Signal to noise ratio of peak flux. (6) Primary beam response at the position of the source. (7) Angular offset from the quasar. $\dagger$: Source has a faint, extended counterpart in K-band ($\sim23$\, AB mag).}
    %\ref{fig:tab2}

\end{deluxetable*}

\begin{deluxetable}{lccc}
\tablenum{3}
\tablecolumns{3}
\tablewidth{0pt}
\tablehead{
  \colhead{log$S_{\nu}$}  & \colhead{N$_{\rm sources}$} & \colhead{$dN/d\rm log S_{\nu}$} & \colhead{$\delta$N} \\
     & & \colhead{($10^3$\,mJy$^{-1}$}\,deg$^{-2}$) & \\
     (1) & (2) & (3) & (4) }

\startdata
    --0.86 & 1 & 45.1 & 46.8\\
    --0.65 & 2 & 19.5 & 14.8\\
    --0.45 & 2 & 9.6 & 7.3\\
    --0.25 & 5 & 20.5 & 10.8\\
    --0.04 & 5 & 20.5 & 10.7\\
    0.16 & - & - & -\\

   \enddata
    \tablecomments{Differential number counts as shown in Figure \ref{fig:cont1}. (1) Lower limit of flux density ($S_{\nu}$) bin. (2) Number of entries per bin (before fidelity and completeness correction). (3) Differential number count (sources per $d$log$S_{\nu}$ per $A_{\rm eff}$), corrected for fidelity and completeness.  (4) Errors on $dN/d\rm log S_{\nu}$.}
\end{deluxetable}

Since the number is nearly indistinguishable from the expectations from blank fields, $N_{\rm exp} = 16.1^{+3.3}_{-6.8}$, there is no statistical overdensity of dust continuum emitters around these quasars.  However, we argue that this is not firm evidence for a lack of clustering or large scale structure around these sources, nor evidence that dust emitters do not exist in protoclusters at these redshifts. The significant result published by \citet{Decarli2017a} shows that [CII] overdensities exist extremely close to the quasar, but this need not be evident in all tracers. There are a number of reasons for a lack of close-by continuum detections:

\textit{ALMA field of view vs. the size of protoclusters.} At $z\approx6$, the ALMA Band 6 primary beam ($25\arcsec$) corresponds only to $\sim$1\,comoving Mpc (cMpc) scales, whereas protoclusters are expected to span $\sim$20\,cMpc along a side, or 2-3\,cMpc in their cores \citep{Overzier2009a, Chiang2017a}, since they have not yet virialized in a compact form. Thus, the small field of view of ALMA may miss the presence of a true overdensity, as dust emitters could sit within a large scale overdensity further away from the quasar. While one might expect to detect an overdensity by observing within 2-3\,cMpc, our 1\,Mpc scales are not sufficient to rule out overdensities (though a detected overdensity would have provided sufficient evidence of a more massive structure). This may be especially true if the quasar itself is emitting ionizing radiation, creating an ionized bubble in the so-called `near-zone' expected to be $\gtrsim1$\,pMpc wide \citep{Fan2006a, Eilers2017a}, which may suppress galaxy formation in the quasar's immediate vicinity. Numerical simulations \citep{Dubois2013a, Costa2014a, Barai2018a} suggest that the most massive quasars at these epochs ($\rm M_{BH} \gtrsim 10^9 M_{\odot}$) may produce significant enough feedback to suppress star formation on hundreds of kpc separation from the quasar, consistent with the present lack of millimeter detections at close angular separations. If this is the case, an excess of dusty star-forming galaxies may not be found on kpc scales near quasars this massive. However, we stress once again that the overdensity of [CII] emitters found by \citet{Decarli2017a} suggests that star-forming galaxies are assembling near the quasars, and it is unlikely that any feedback process would eliminate the dust emission but not the gas emission. It is still possible, however, that true overdensities of dust emitters would be traced only on much larger scales (several comoving Mpc). To test this hypothesis, observations targeting a much wider field around quasars are required.

\textit{Cosmic Variance.} %The small size of the primary beam and the small total area of this survey cause the additional issue of cosmic variance to arise. 
The average ALMA primary beam ($\sim25\,\arcsec$) is not only much smaller than the size of a typical protocluster, but is also significantly smaller than the physical scale of expected density variations in the $z\sim6$ Universe \citep[30-80\,Mpc or 0.25-0.5\,deg; e.g.,][]{DAloisio2018a}. Despite the fact that quasar sightlines are assumed to be biased toward overdense regions, both cosmic variance and Poisson noise contribute to large fluctuations in field-to-field number counts. According to estimates of mass overdensities from protocluster simulations at $z\sim5$ \citep[e.g.,][]{Chiang2013a}, $\delta_{gal}$ may vary by a factor of $2-3$ between protoclusters, depending on the mass scale of the protocluster and the physical volume probed. This variance rises when the field of view shrinks, so with low numbers of quasars whose individual fields of view are small, it is entirely possible that the field-to-field variation would not amount to an aggregate statistical overdensity due to cosmic variance alone.
To combat this, a larger sample size of random quasar sightlines or wide-field mosaicing around one quasar is necessary.
    
To improve upon the small number statistics, we might consider ALMA observations that reach a fainter flux limit than the one in this study, since the slope of the 1.2\,mm number counts predicts that sources with $S_{\nu}<0.1$\,mJy will be more common by an order of magnitude than our current limit. However, because of the strong negative K-correction for sub/millimeter flux, this method would likely introduce many more faint foreground contaminants. Additionally, the faint-end slope of the IR luminosity function is much shallower than that of the UV luminosity function as probed by Ly$\alpha$ \citep{Casey2014a, Finkelstein2015a}; therefore, increased depth may not reveal an overdensity of IR sources currently below the detection threshold, unlike the large effect of depth on UV/optical surveys.

\textit{Density of foreground sources.}
The largest effect on this study is likely the large comoving depth of this search, due to the flat redshift slelection of 1.2\,mm dust observations. The very negative K-correction of the sub/millimeter regime means sources do not fade with increasing redshift from $z\sim1-10$ \citep[see Fig. 3;][]{Casey2014a}, so compared with the narrow redshift ranges for LAE and [CII] overdensity searches, we are probing an enormous line of sight volume with potential for many millimeter detections. If we are averaging over Gpc scales along the line of sight, it is likely that clusters and voids at various redshifts will cancel out any potential overdensity signal near the targeted quasars. This is especially true as the cosmic IR luminosity function suggests that dusty IR sources are much more common at redshifts $1<z<3$ \citep{LeFloch2005a, Gruppioni2013a}, so at these wavelengths there are probably many more galaxies in the foreground than at the redshift of the quasars. 
    
    Without spectroscopic redshifts, or at the least a strong characterization of the luminosity functions at $z<5$, our current data do not achieve high enough contrast against the foreground population. Future continuum studies would be aided by strong observational constraints of the number of expected foreground sources in continuum, or by obtaining spectroscopic redshifts before looking for dusty counterparts.

\citet{Chiang2013a} uses semi-analytical models of protocluster evolution to predict their overdensity profiles out to $z=5$ in cubes of (15 cMpc)$^3$. The profile with the closest analog to this work is for star-forming galaxies with SFR $>1 \rm  M_{\odot} \rm yr^{-1}$ (although the sources here are likely to be much rarer SMG-type galaxies), and assuming the conditions at $z=5$ are applicable to this work, we use this to predict $\delta_{\rm gal}$ in observations. While the observed overdensity will appear less significant in practice due to the background of sub/millimeter sources, we expect the overdensity to be much higher closer to the density peak, so we may still be able to detect a significant $\delta_{\rm gal}$ given these models. 

As a first step, we convert from $\delta_{\rm gal}$ in \citet{Chiang2013a} to a 2D projection in an aperture the size of the ALMA Band 6 primary beam. To estimate the expected number of sources at this redshift, we would ideally use the infrared luminosity function (IRLF) at $z=6$, but as this has not yet been measured \citep{Casey2018a}, we assume to first order an expected volume density of sources matching the $z=6$ LBG luminosity function found in \citet{Finkelstein2016a}. After converting the 3D $\delta_{\rm gal}$ from \citet{Chiang2013a}, we expect $N_{\rm gal} \approx 0.13$ associated sources per primary beam, if the quasar resides at the density peak. Thus, this translates to an estimated 4.5 sources associated with 35 $z=6$ quasars, or a 2D $\delta_{\rm gal} \approx 0.3$ when including the expected millimeter foreground sources based on current blank field number counts. This highlights the difficulty in finding true overdensities in millimeter continuum at high redshift, since $\delta_{\rm gal}$ is so easily washed out and difficult to constrain without a firm grasp on the IR field population at comparable epochs. Additionally, since neither the $z=6$ IRLF nor the galaxy bias for SMGs has been observationally constrained, we need more data to make further predictions. However, if semi-analytical models can even approximately estimate SMG overdensities in protoclusters centered on quasars, observational evidence is within reach with a larger sample size.\\

In summary, 15 candidate detections at $\geq 5\sigma$ in 11 out of 35 ALMA maps centered on $z\sim6$ quasars does not represent a statistically significant flux excess when compared with blank field number counts at 1.2\,mm, with a measured $\delta_{\rm gal} = -0.07\pm0.56$. The discrepancy between this result and the [CII] overdensity found by \citet{Decarli2017a} is likely due to the large comoving volume probed by our millimeter continuum study, since we cannot confirm that the detected sources are companions to the quasars unless they had a prior [CII] redshift. We can motivate future tests of the hypothesis that SMGs can exist in protoclusters centered around quasars through much deeper observations, or by substantially widening the field of view through mosaic observations. Deeper observations may reveal fainter [CII] sources as well, enabling us to constrain the redshifts of the sources found in the fields. Another follow-up method of study could be going to shorter wavelengths and searching for LAEs with \textit{HST} grism observations, or searching for counterparts in stellar light using \textit{JWST}, which would aid in spectroscopic confirmation. We do not take this result to be evidence that protoclusters at high redshift lack dusty star-forming galaxies, as further observations are required to confirm this result. \\

\acknowledgements

We thank the anonymous referee for their insightful comments which helped to improve this paper. JBC and CMC thank the College of Natural Sciences at the University of Texas at Austin for support in addition to NSF grant AST-1714528. D.R. acknowledges support from the national Science Foundation under grant number AST-1614213. B.V. and F.W. acknowledge funding through ERC grants ``Cosmic Dawn" and ``Cosmic Gas."

This paper makes use of the following ALMA data: ADS/JAO.ALMA:  2011.0.00206.S (PI: R. Wang), 2011.0.00243.S (PI: C. Willott), 2012.1.00882.S (PI: B. Venemans), 2013.1.00273.S (PI: B. Venemans), 2015.1.00997.S (PI: R. Maiolino), and 2015.01115.S (PI: F. Walter). ALMA is a partnership of ESO (representing its member states), NSF (USA) and NINS (Japan), together with NRC (Canada) and NSC and ASIAA (Taiwan) and KASI (Republic of Korea), in cooperation with the Republic of Chile. The Joint ALMA Observatory is operated by ESO, AUI/NRAO and NAOJ.

The National Radio Astronomy Observatory is a facility of the National Science Foundation operated under cooperative agreement by Associated Universities, Inc.

%\clearpage
\nocite{*}
\bibliography{main}

\clearpage

\appendix 
\section{}
\subsection{Quasar Maps with No Additional Detections}

\begin{figure*}[h]
\begin{center}
    \includegraphics[width=0.49\columnwidth]{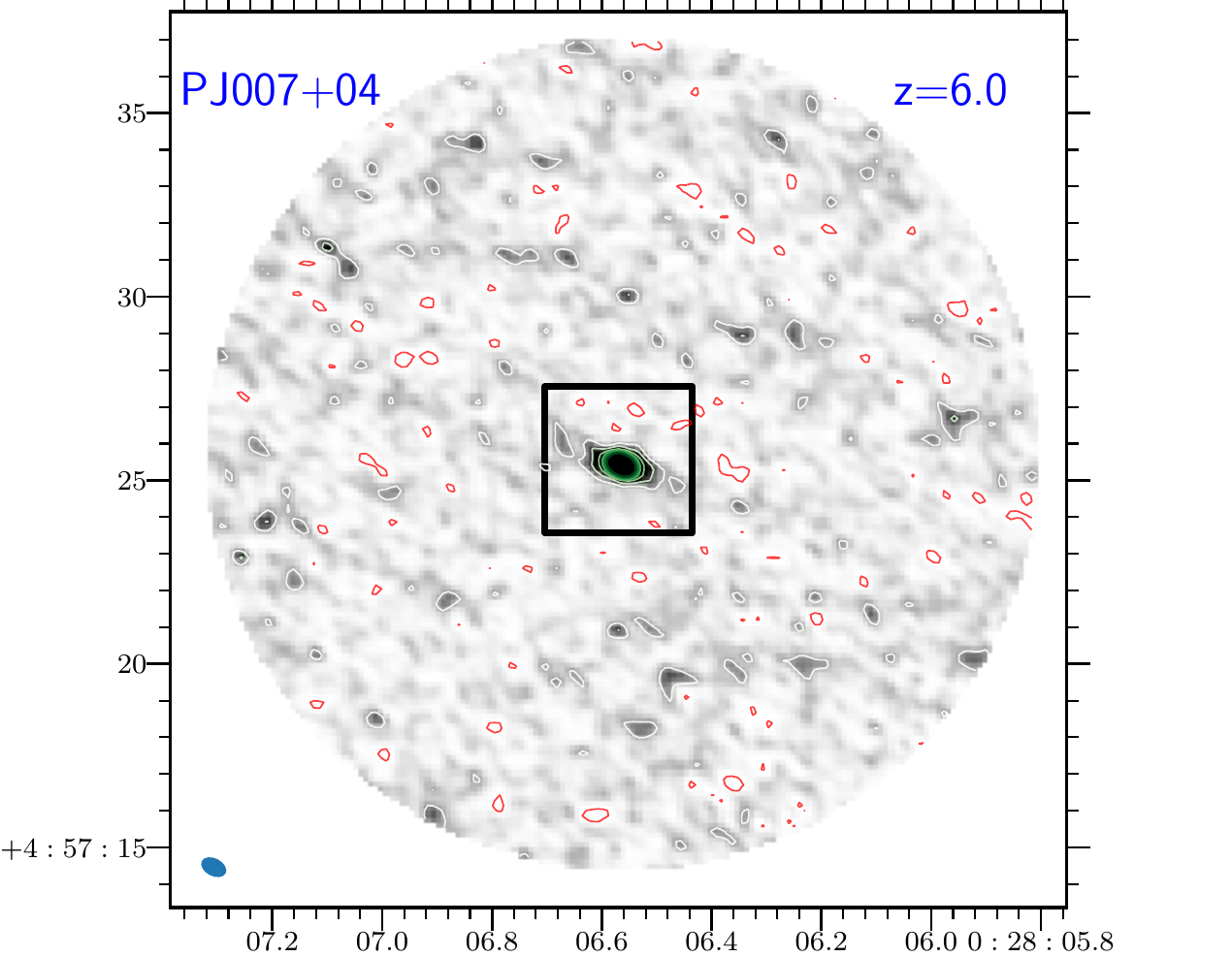}
    \includegraphics[width=0.49\columnwidth]{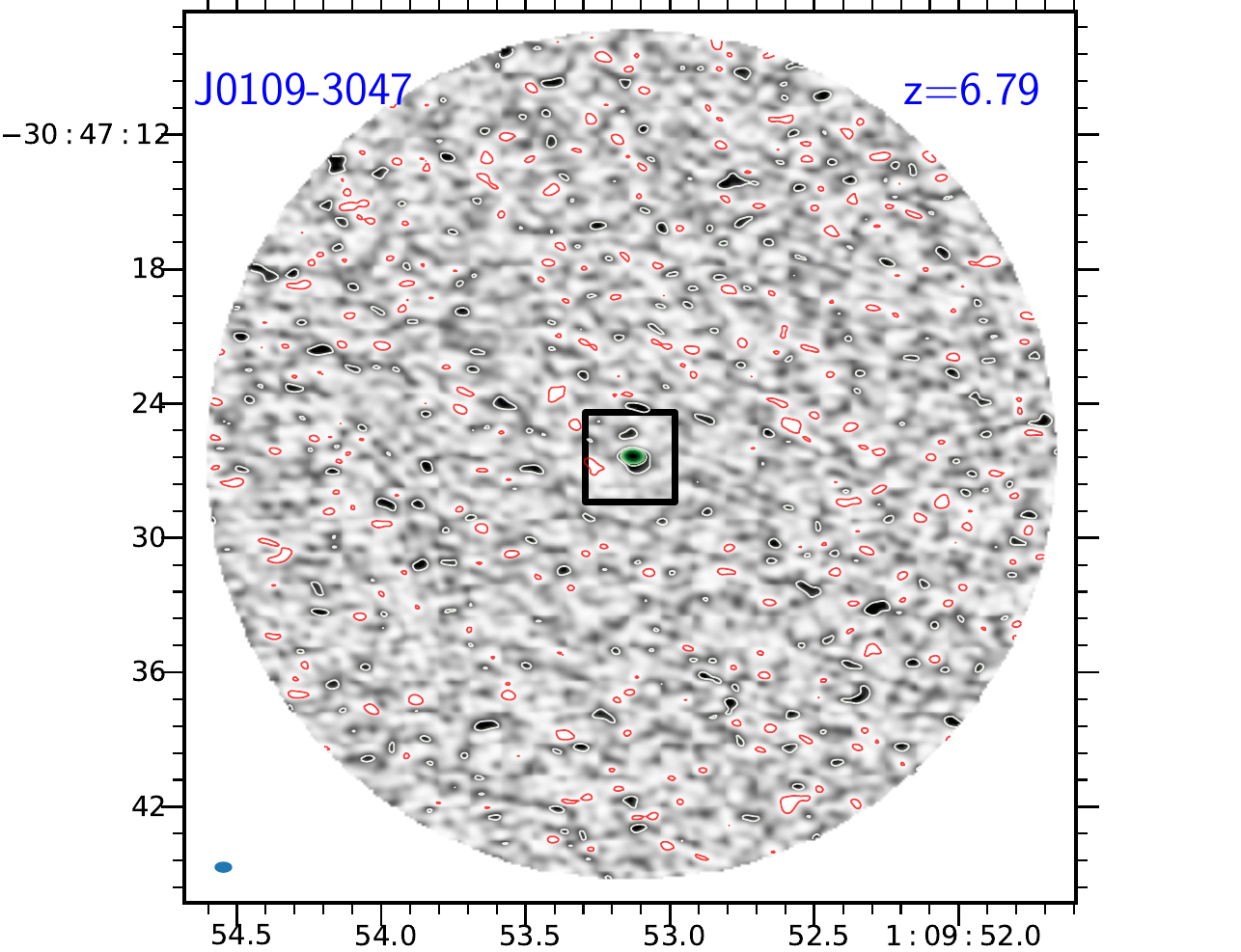}
    \includegraphics[width=0.49\columnwidth]{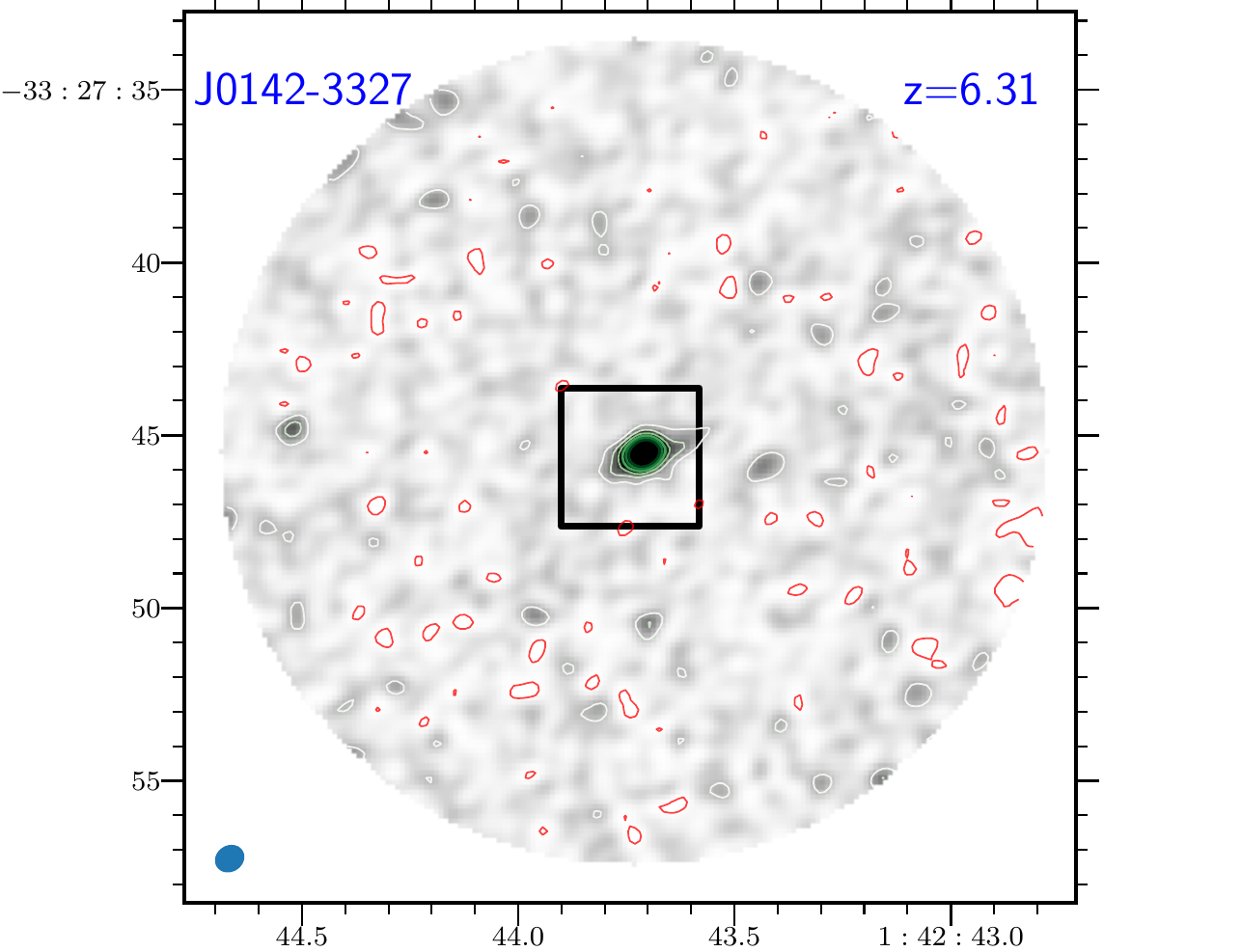}
    \includegraphics[width=0.49\columnwidth]{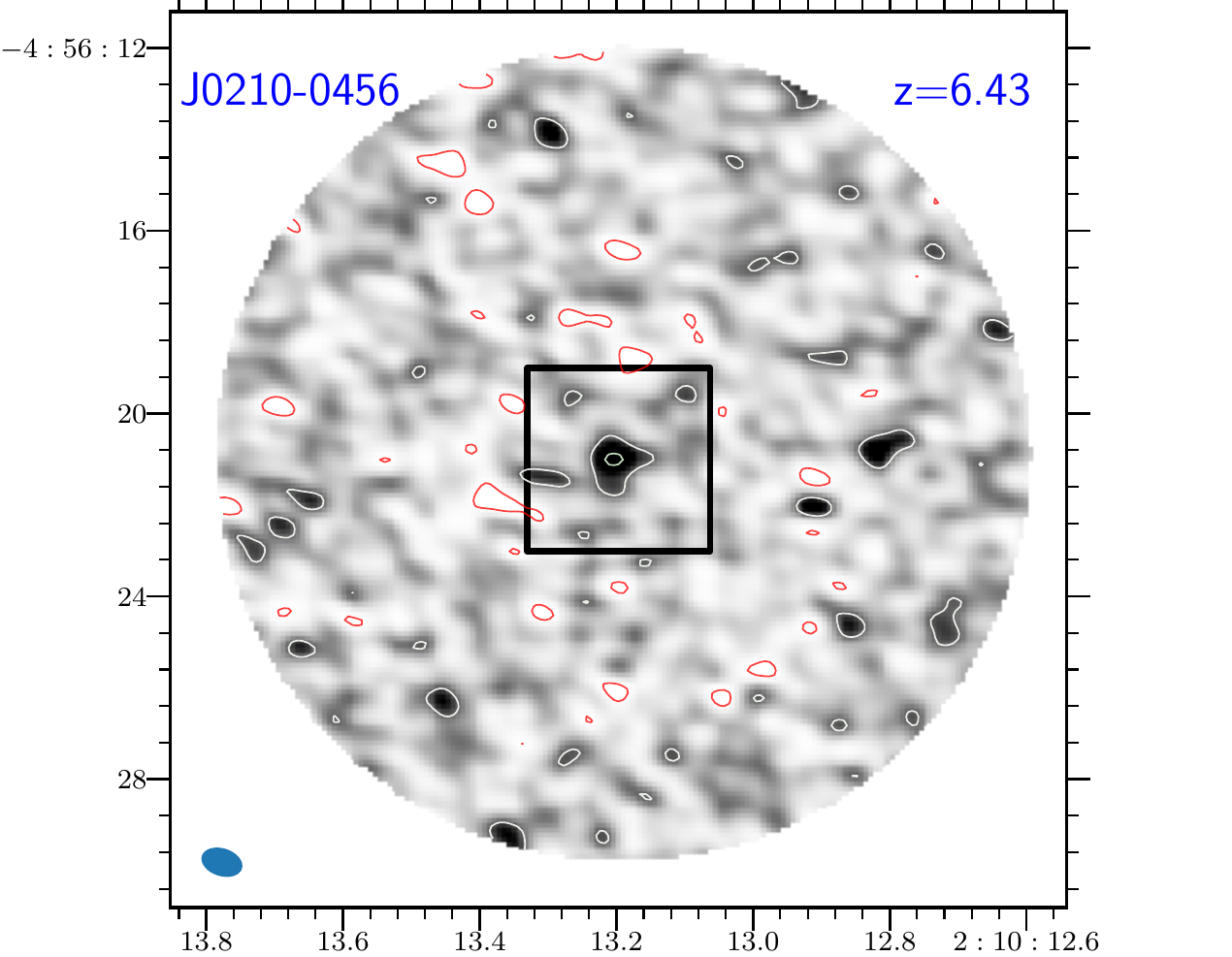}
    \includegraphics[width=0.49\columnwidth]{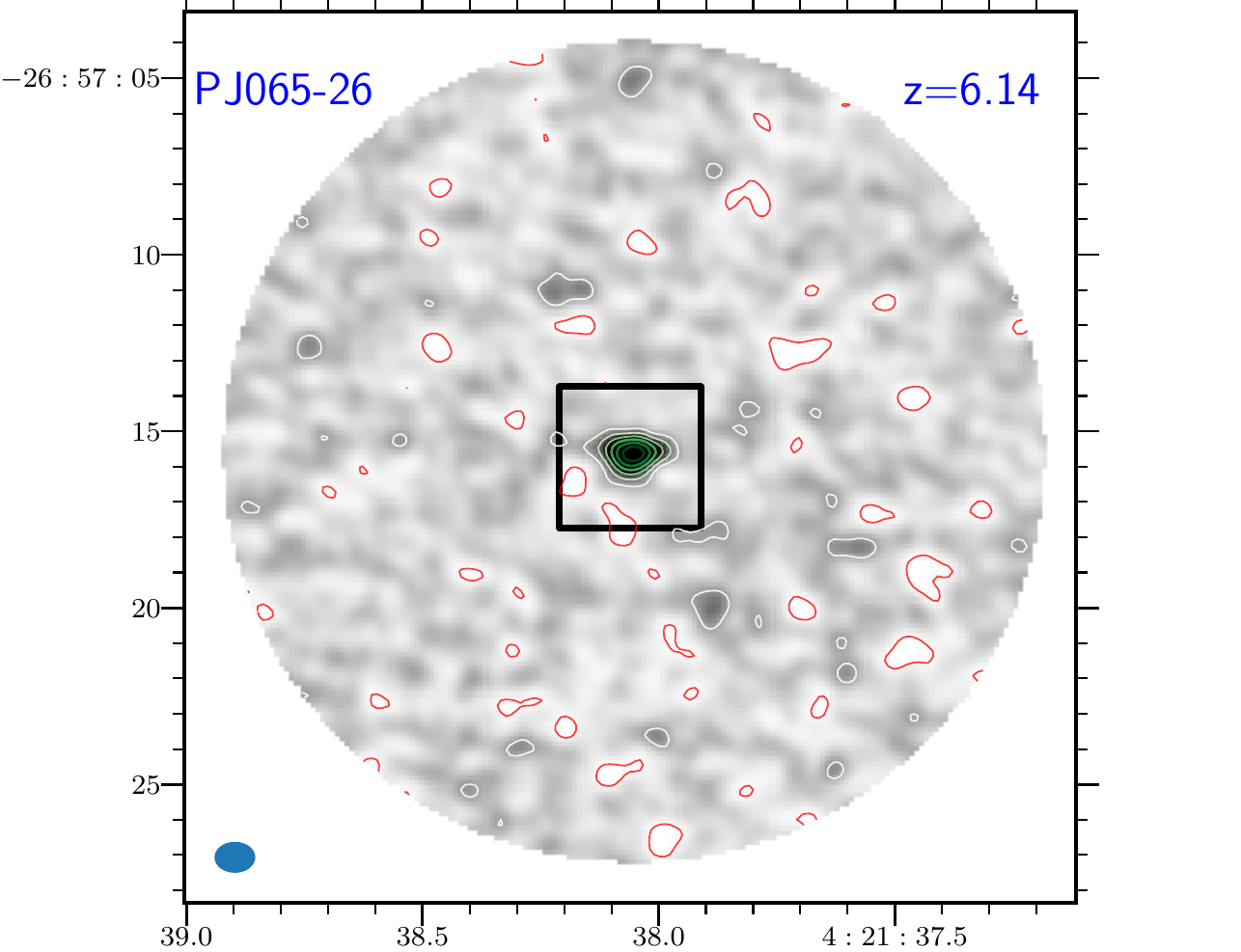}
    \includegraphics[width=0.49\columnwidth]{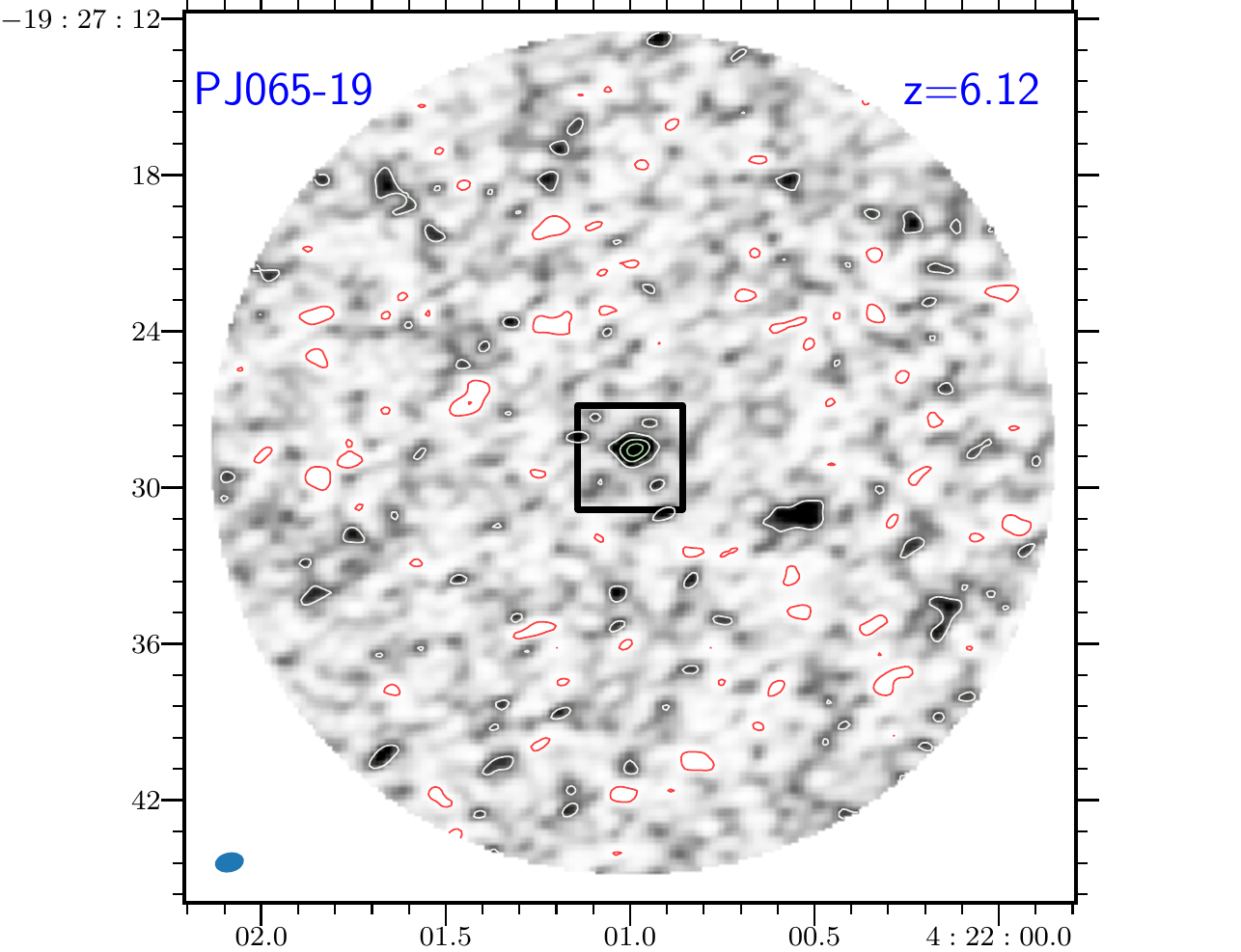}

\end{center}
\end{figure*}

\begin{figure*}[h]
\begin{center}
    \includegraphics[width=0.49\columnwidth]{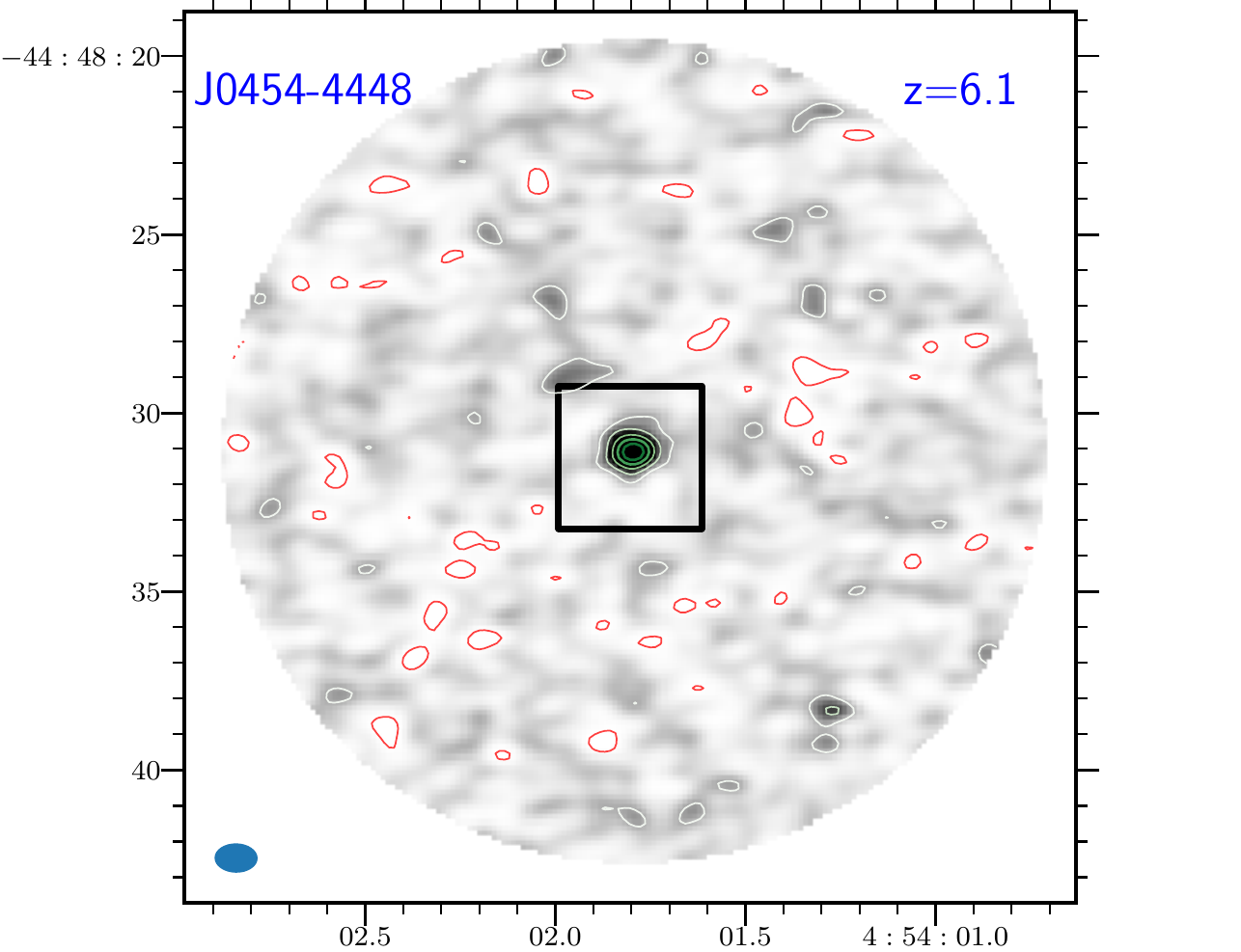}
    \includegraphics[width=0.49\columnwidth]{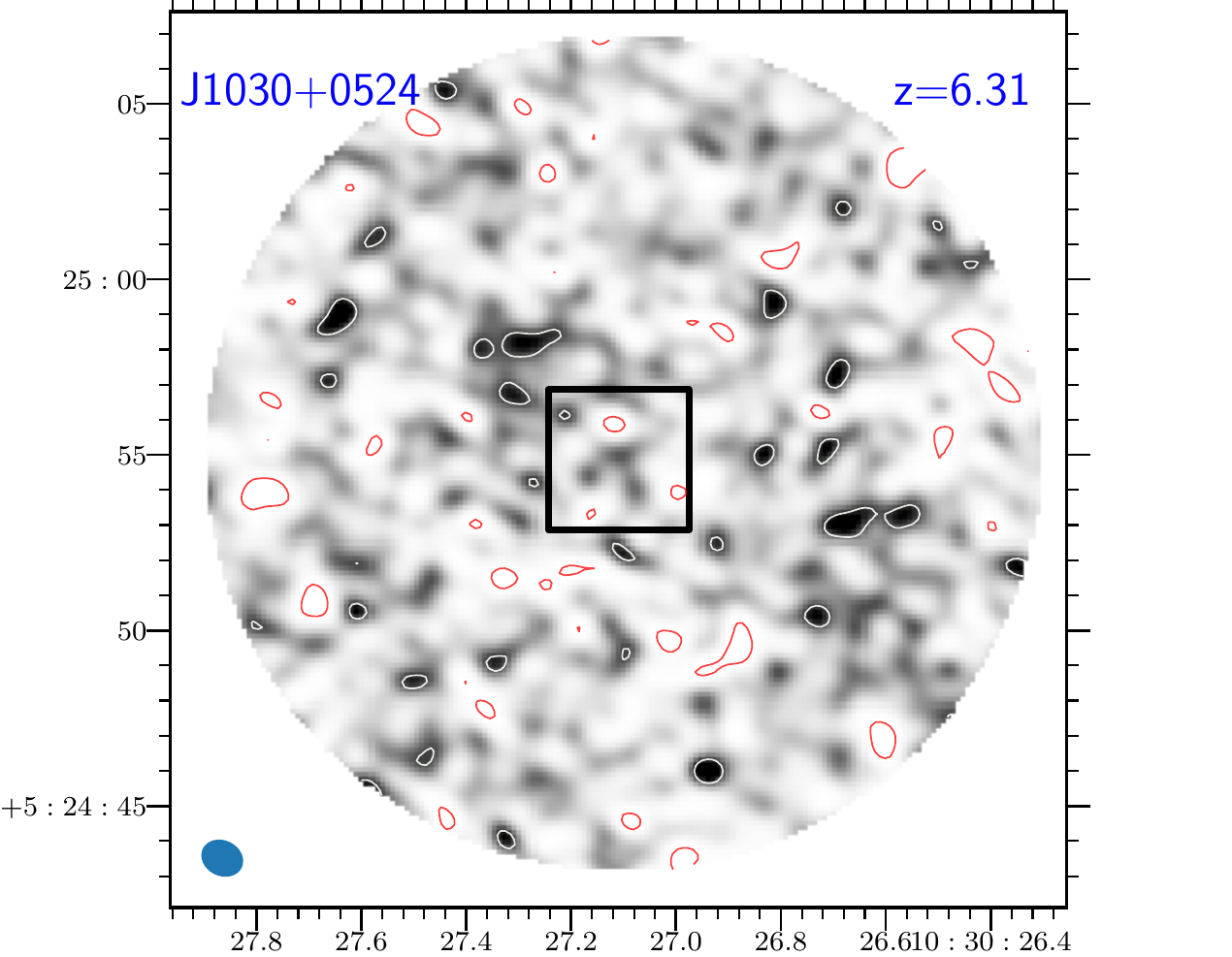}
    \includegraphics[width=0.49\columnwidth]{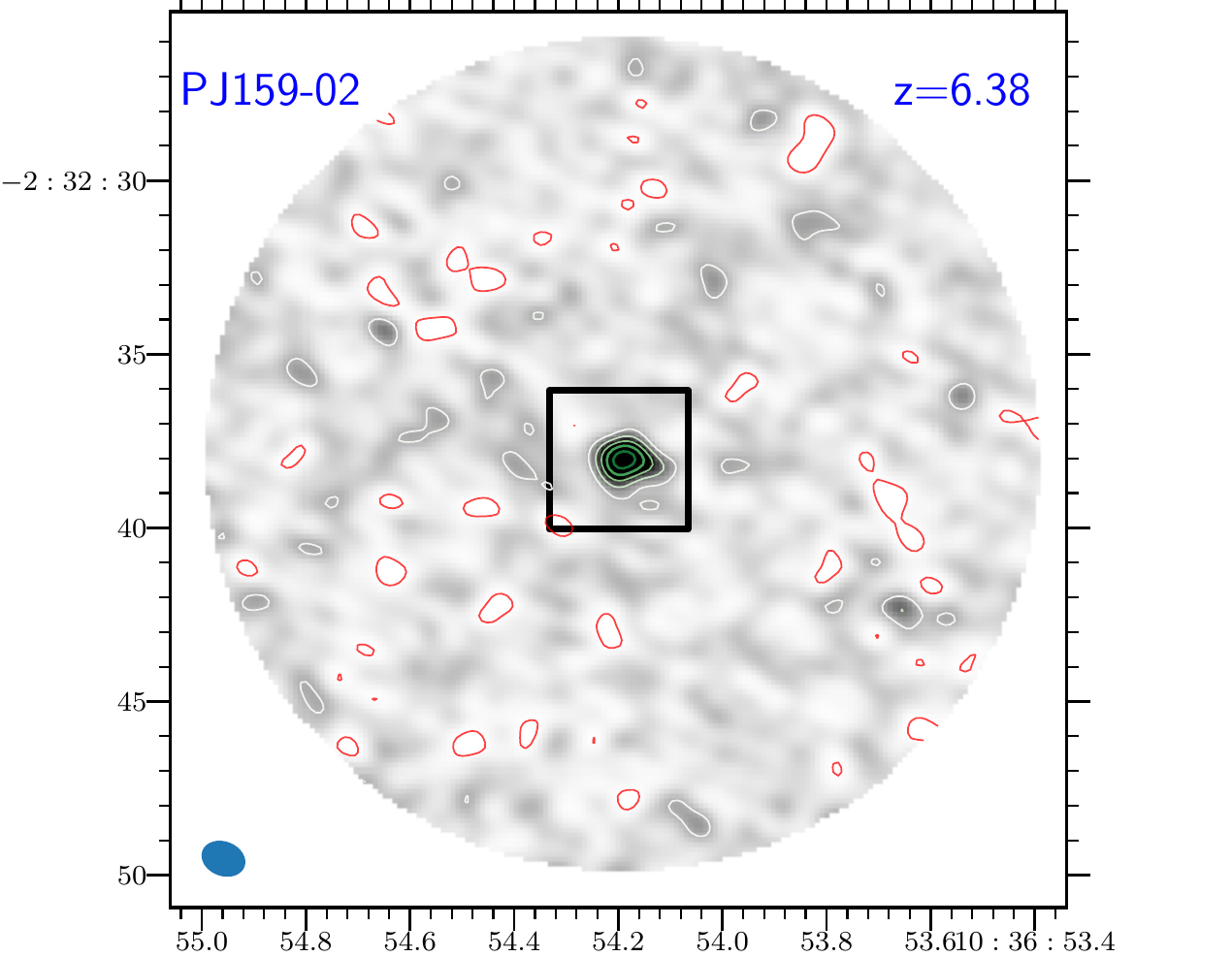}
    \includegraphics[width=0.49\columnwidth]{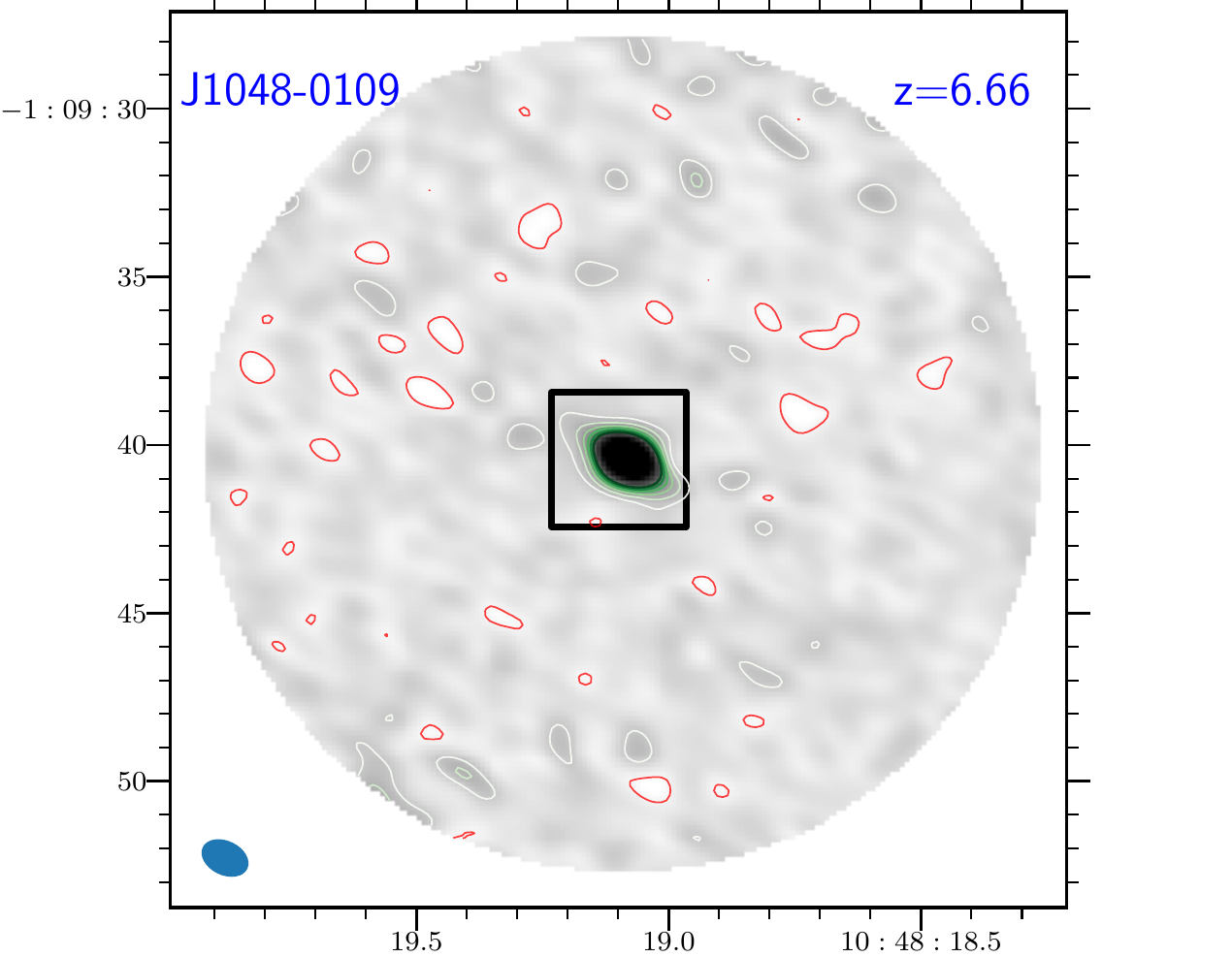}
    \includegraphics[width=0.49\columnwidth]{PJ167-13_nomask_contour.pdf}
    \includegraphics[width=0.49\columnwidth]{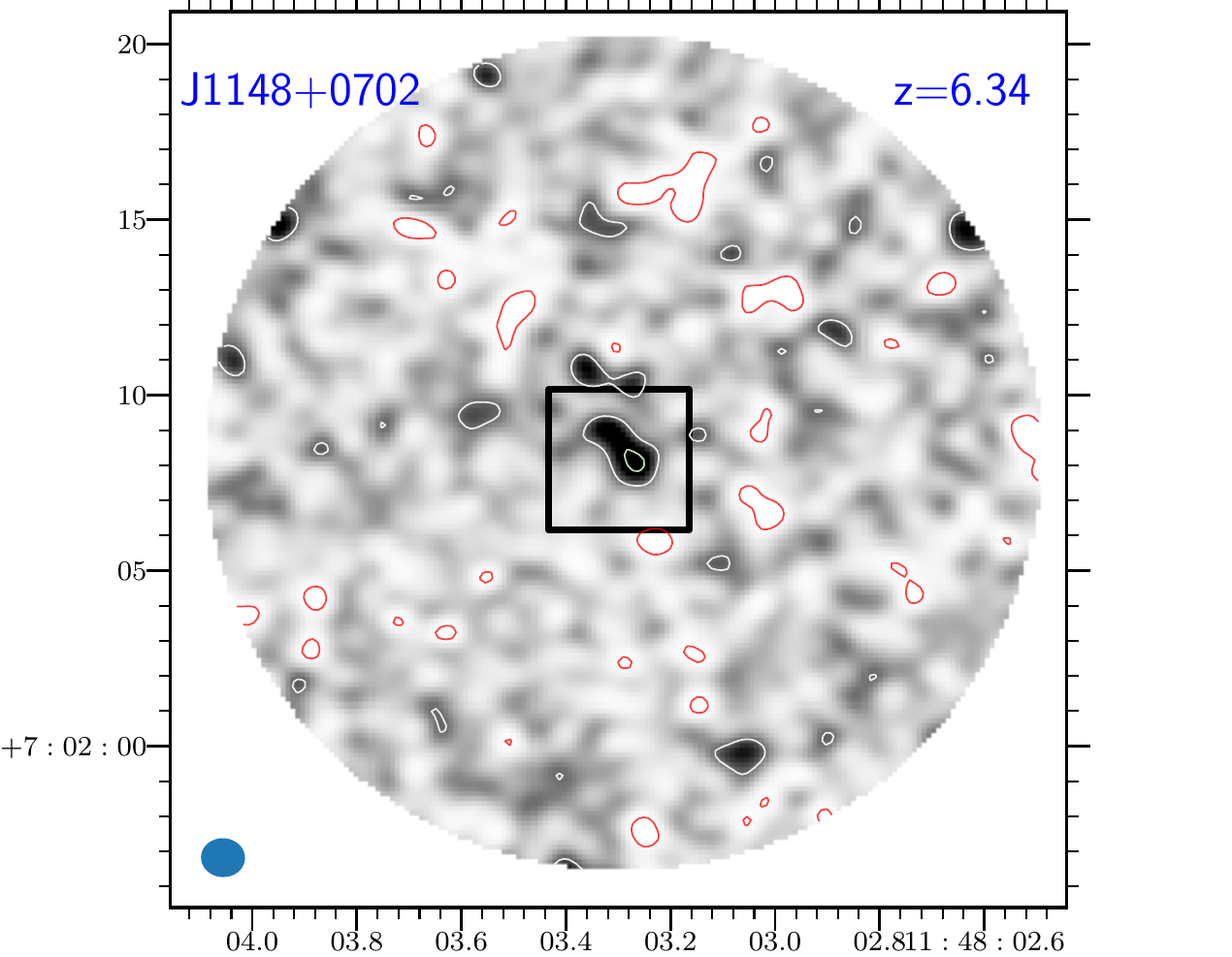}
\end{center}
\end{figure*}

\begin{figure*}[h]
\begin{center}
     \includegraphics[width=0.49\columnwidth]{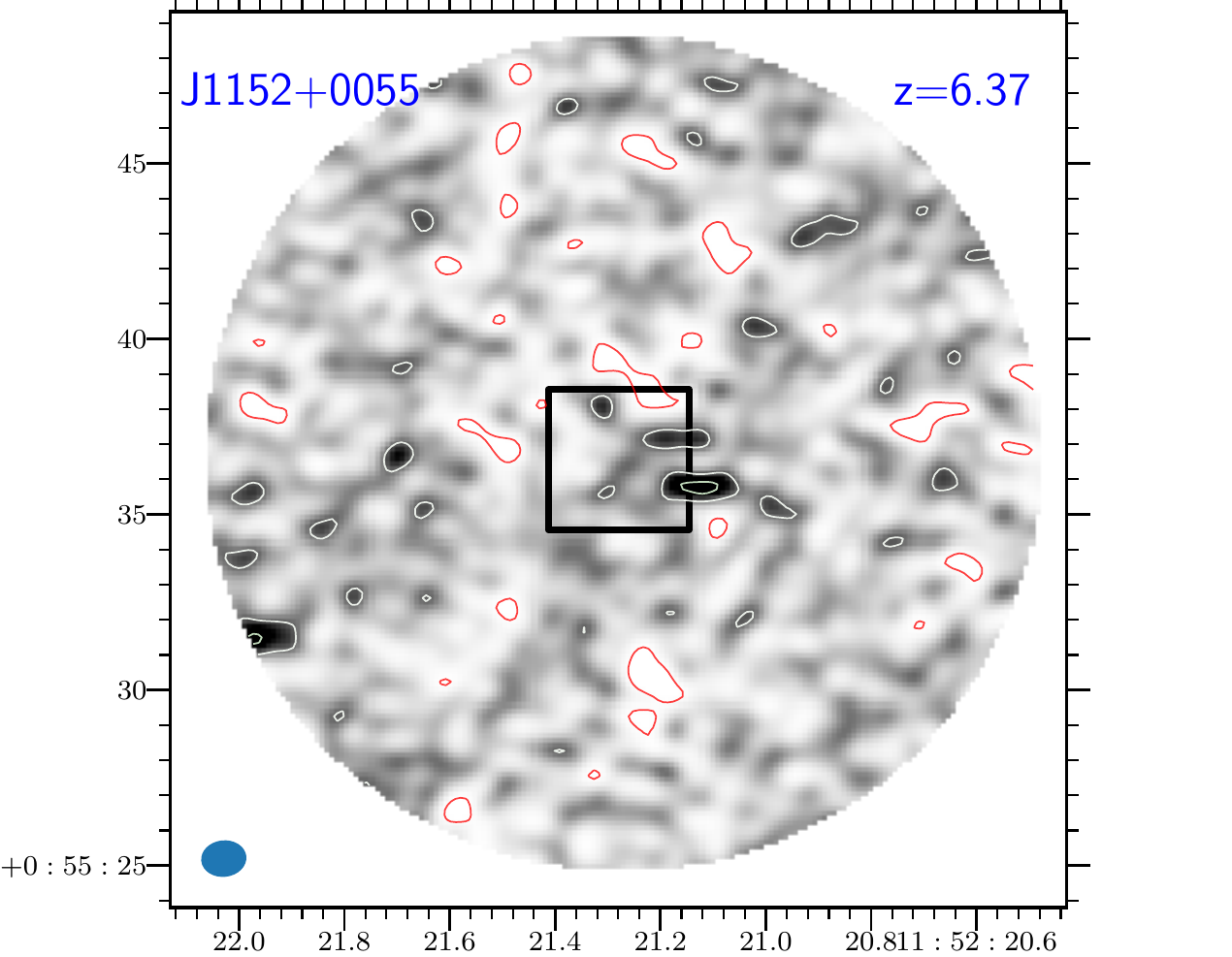}
     \includegraphics[width=0.49\columnwidth]{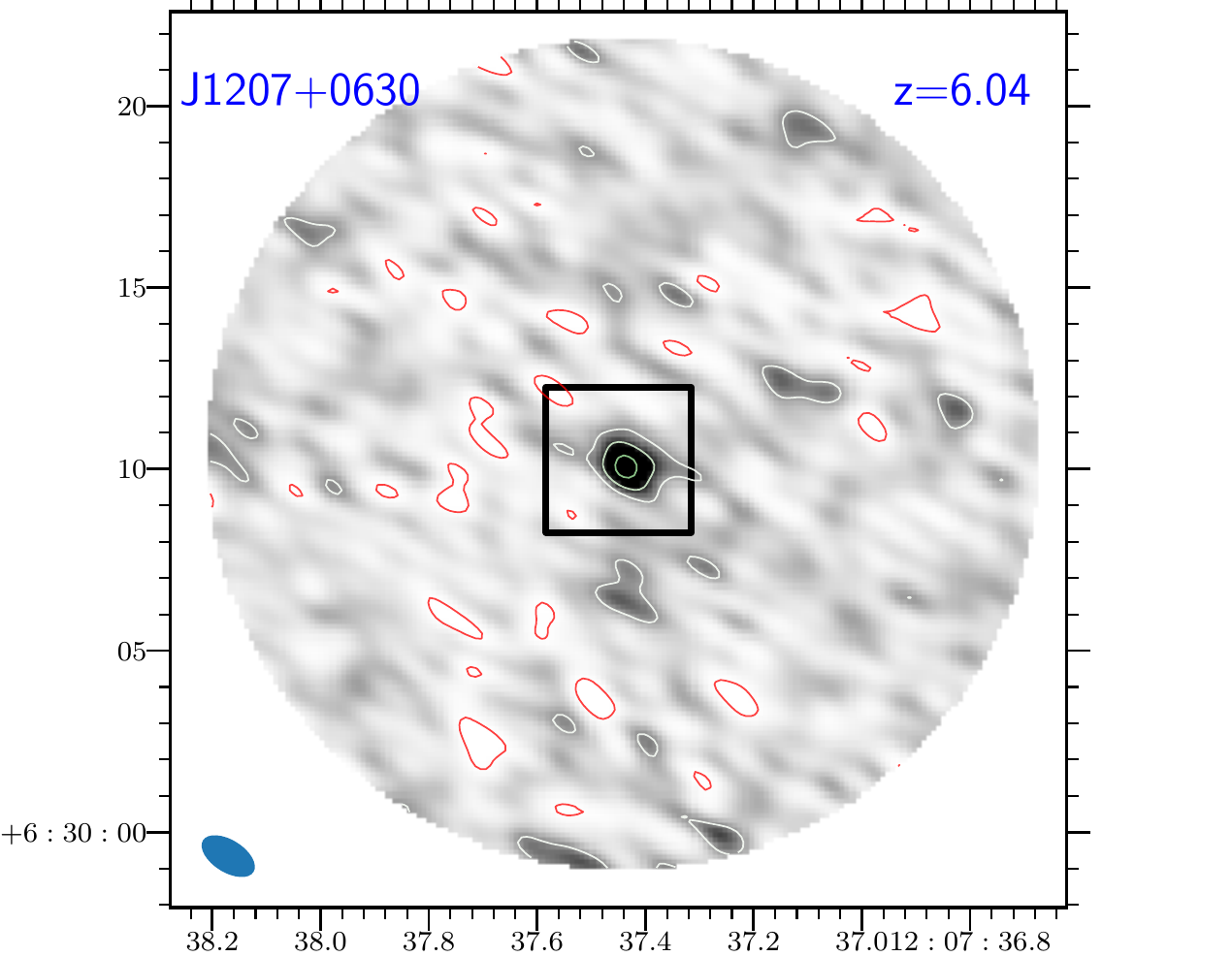}
    \includegraphics[width=0.49\columnwidth]{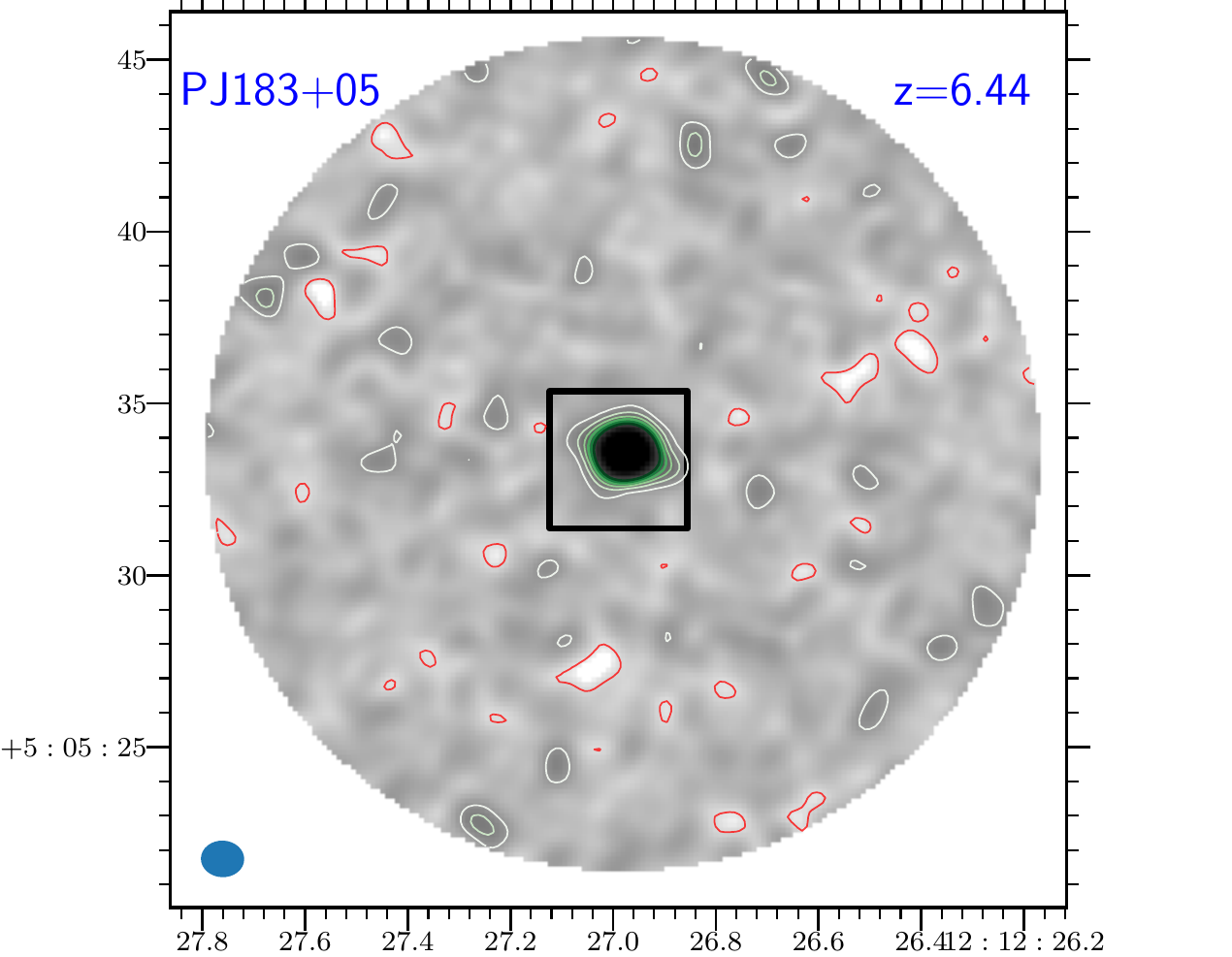}
    \includegraphics[width=0.49\columnwidth]{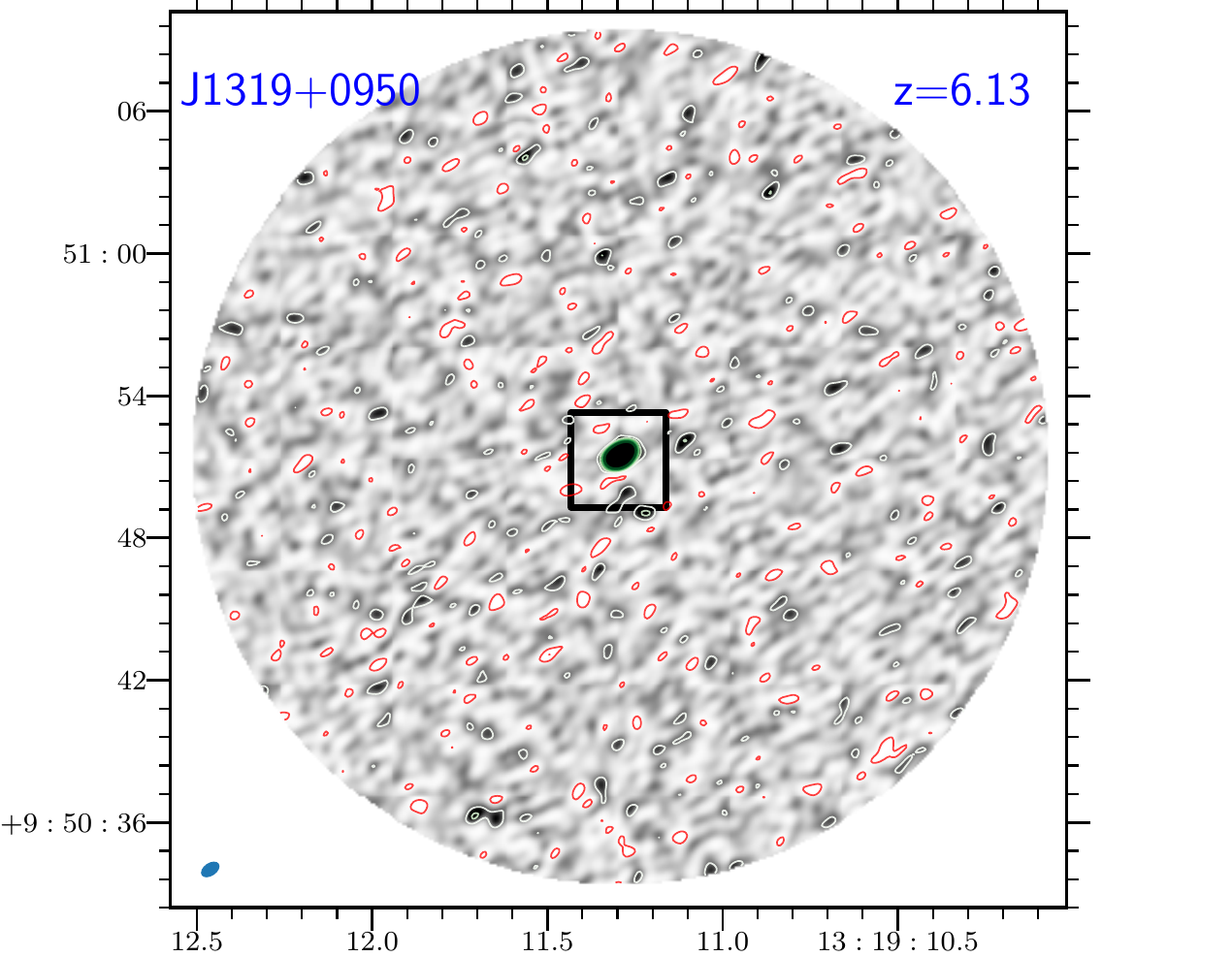}
    \includegraphics[width=0.49\columnwidth]{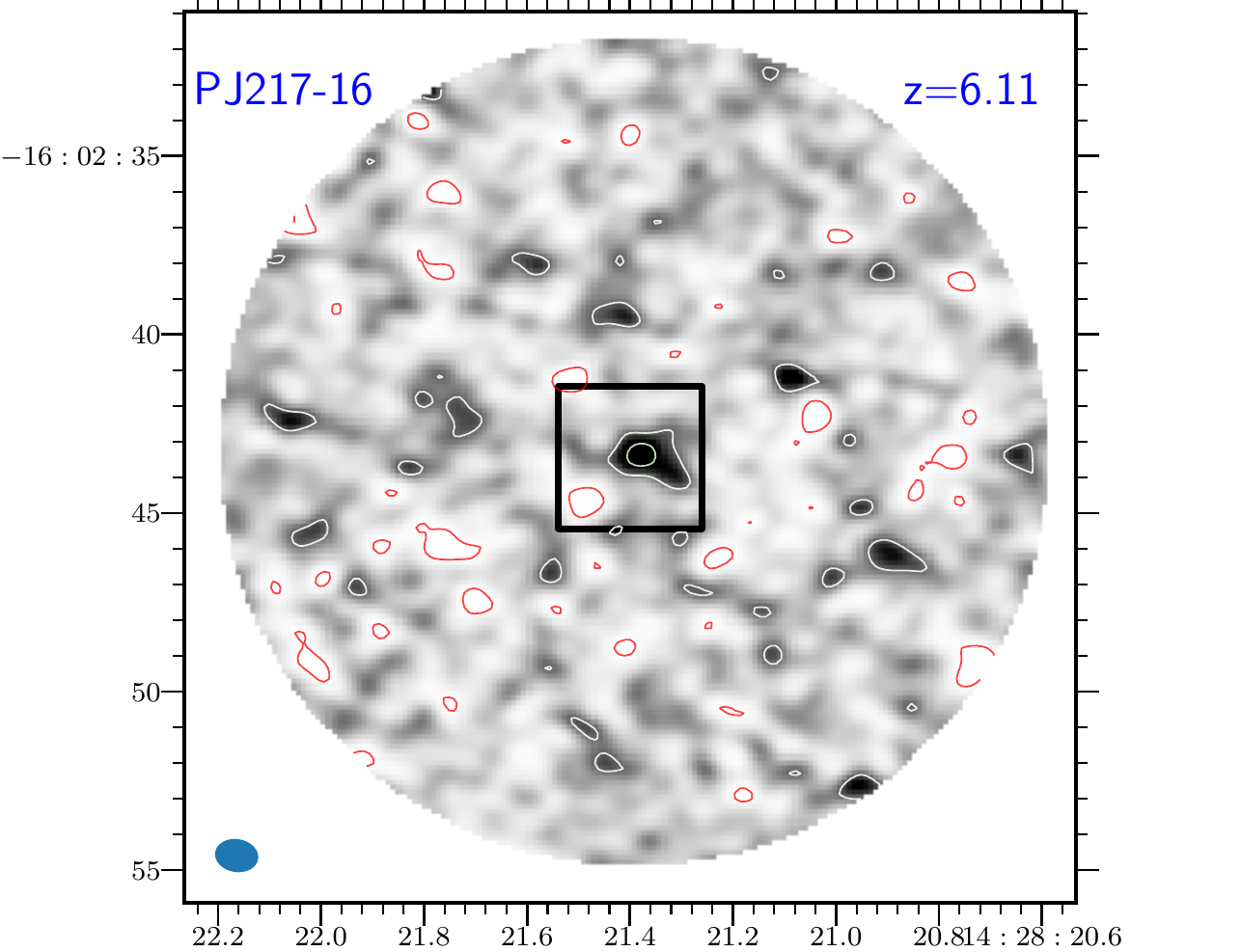}
    \includegraphics[width=0.49\columnwidth]{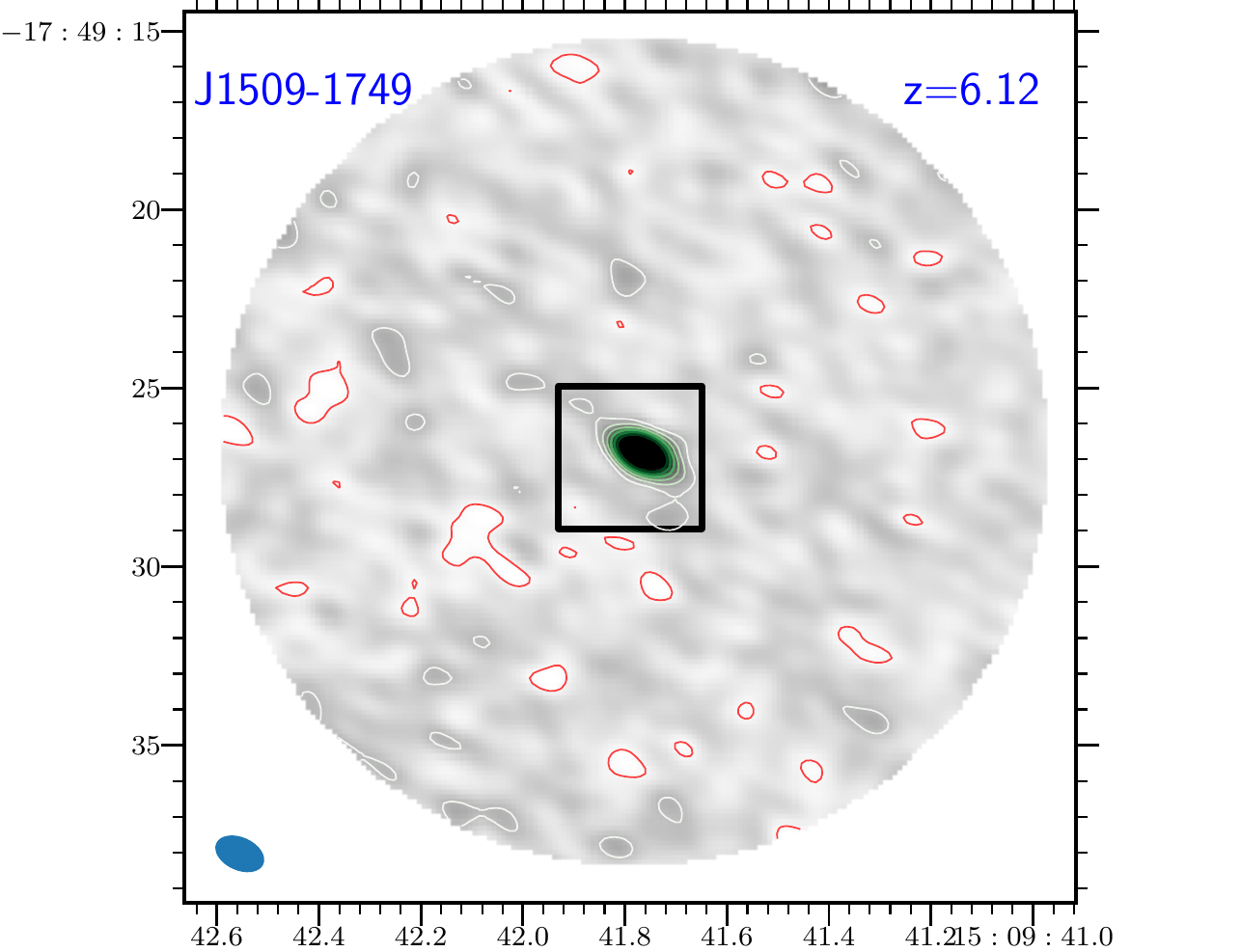}
    
\end{center}
\end{figure*}

\begin{figure*}[h]
\begin{center}
    \includegraphics[width=0.49\columnwidth]{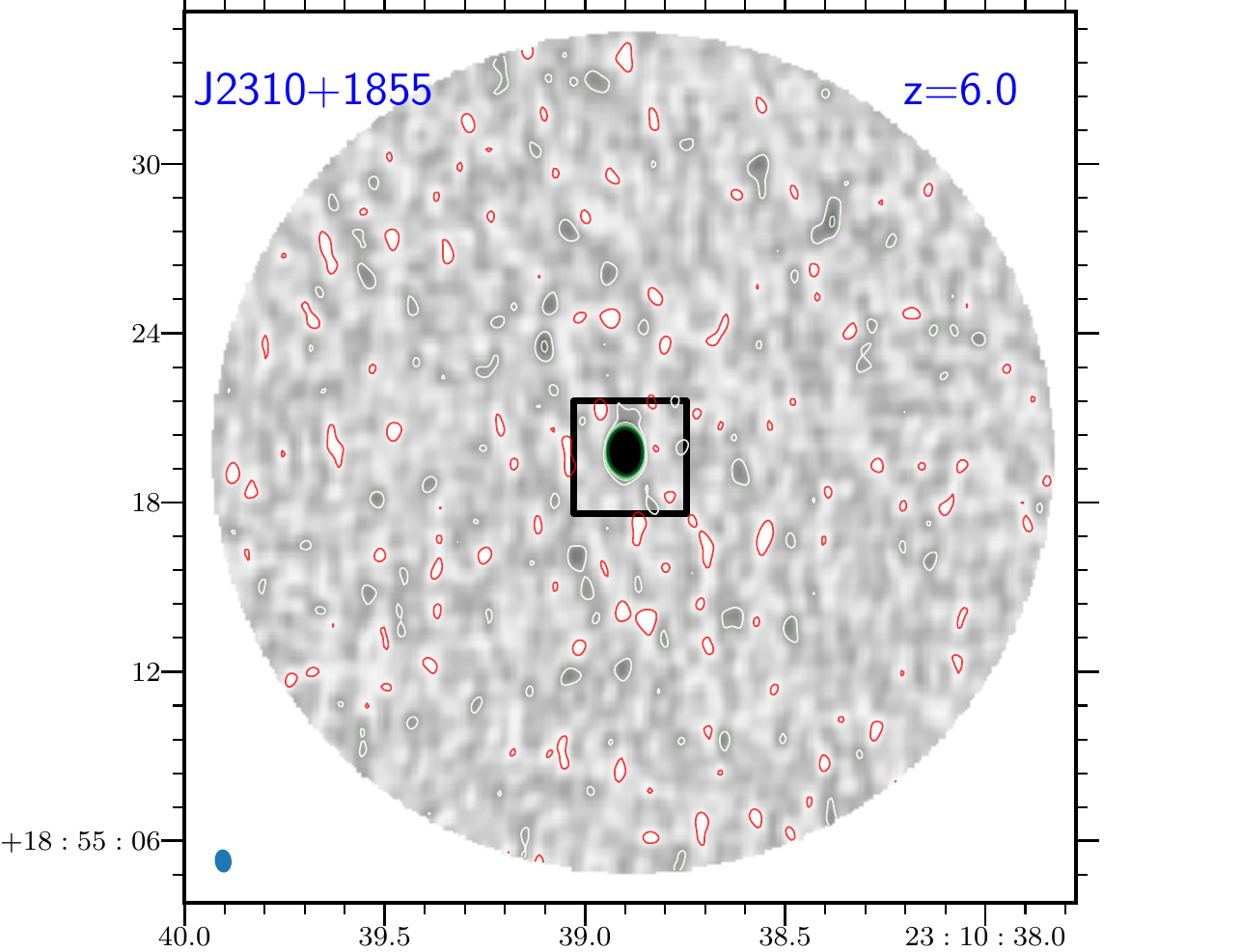}
    \includegraphics[width=0.49\columnwidth]{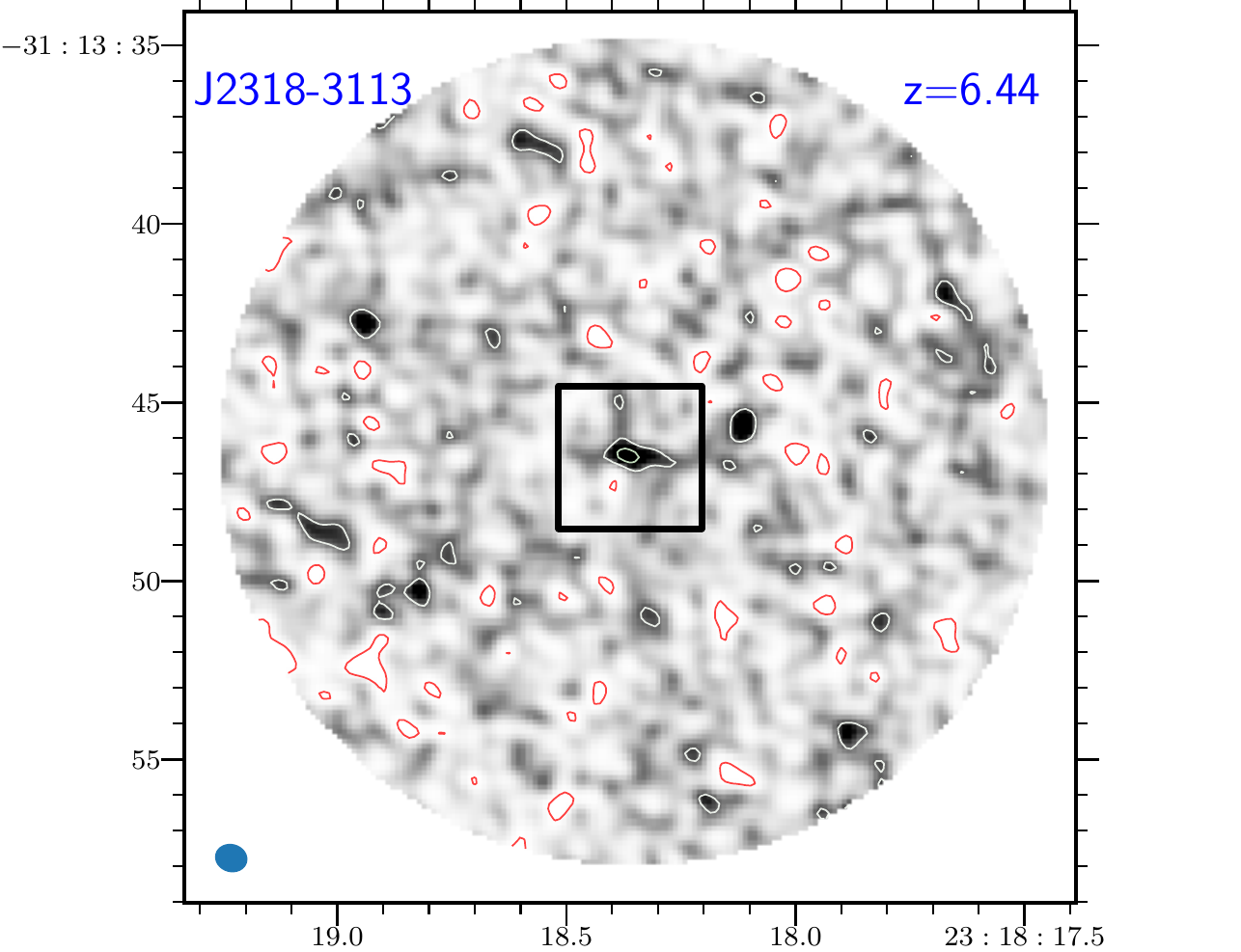}
    \includegraphics[width=0.49\columnwidth]{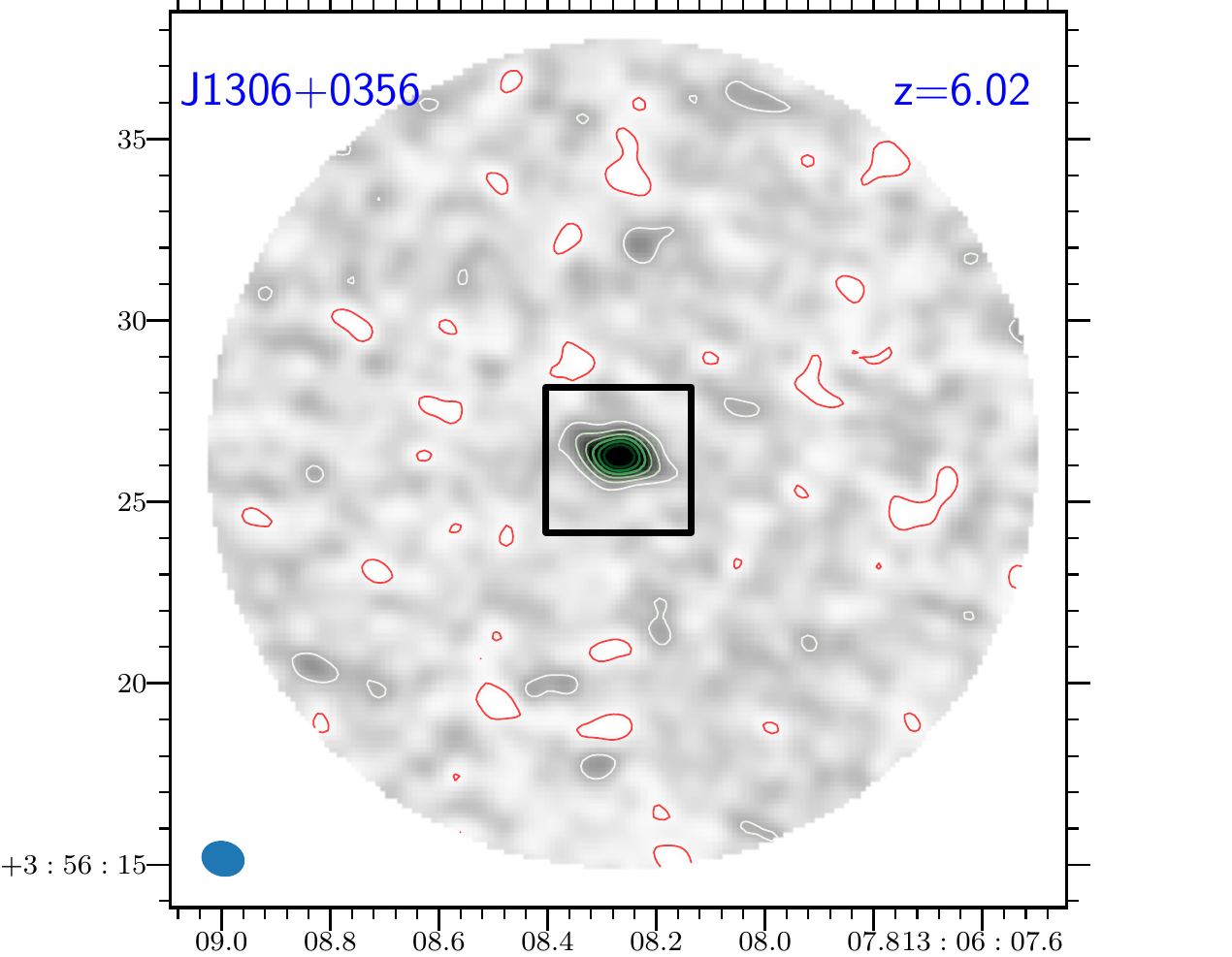}
    \includegraphics[width=0.49\columnwidth]{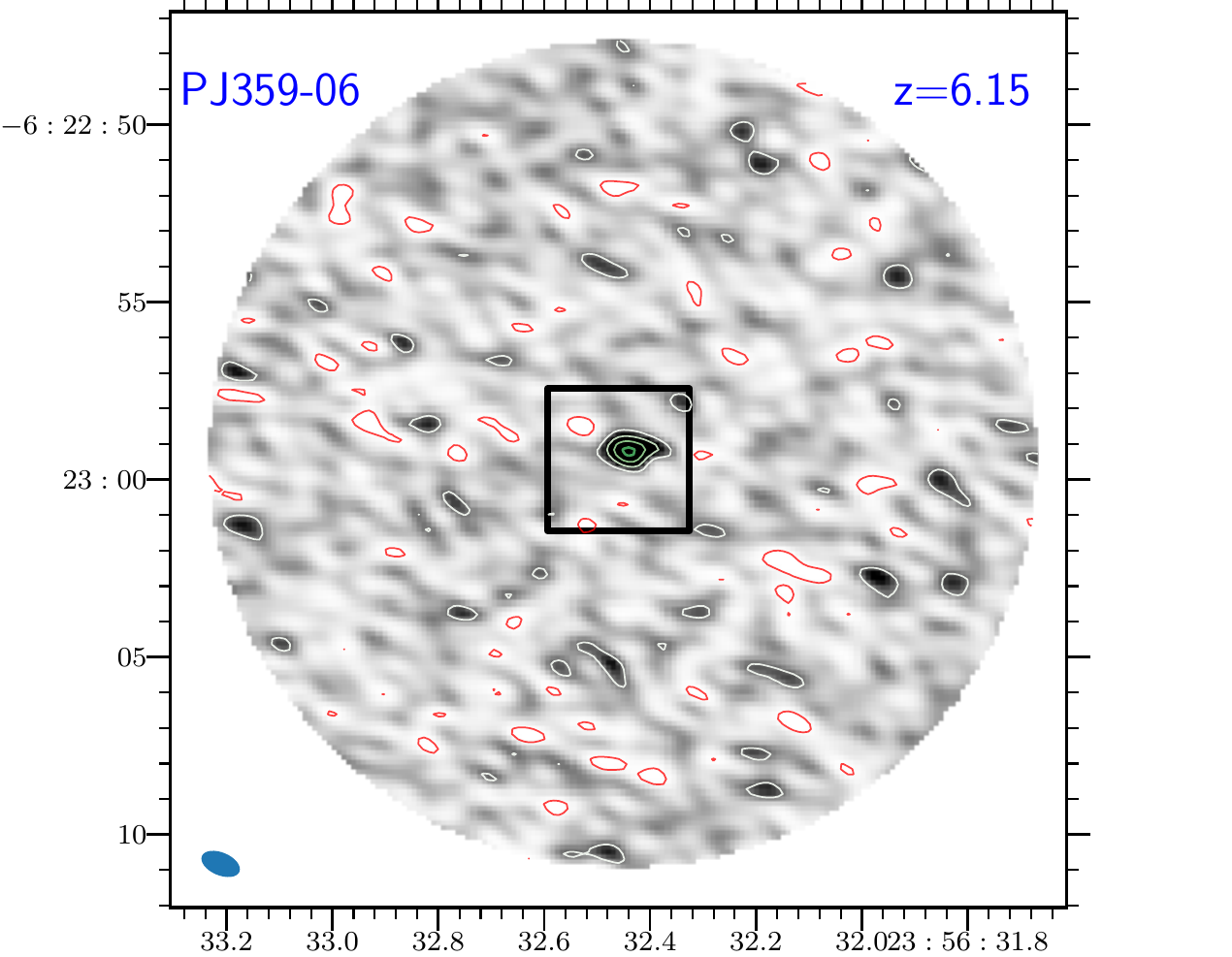}
\end{center}
\end{figure*}

\clearpage

\end{document}